 \documentclass[twocolumn]{aastex61}

\usepackage{amsmath}

\received{April 20, 2017}
\revised{TBD}
\accepted{TBD}
\submitjournal{ApJ}
	
\shorttitle{Model independent exoplanet transit spectroscopy}
\shortauthors{Aronson et al.}

\begin{document}

\title{Model-independent exoplanet transit spectroscopy}

\correspondingauthor{Erik Aronson}
\email{erik.aronson@physics.uu.se}

\author{Erik Aronson}
\affiliation{Observational Astronomy, Department of Physics and Astronomy, Uppsala University, Box 516, SE-751 20 Uppsala, Sweden}

\author{Nikolai Piskunov}
\affiliation{Observational Astronomy, Department of Physics and Astronomy, Uppsala University, Box 516, SE-751 20 Uppsala, Sweden}

\begin{abstract}
We propose a new data analysis method for obtaining transmission spectra of exoplanet atmospheres and brightness variation across the stellar disk from transit observations. The new method is capable of recovering exoplanet atmosphere absorption spectra and stellar specific intensities without relying on theoretical models of stars and planets. We fit both stellar specific intensity and planetary radius simultaneously directly to transit light curves. This allows stellar models to be removed from the data analysis. Furthermore, we use a data-quality weighted filtering technique to achieve an optimal trade-off between spectral resolution and reconstruction fidelity homogenising the signal to noise ratio across the wavelength range. Such approach is more efficient than conventional data binning onto a low resolution wavelength grid. We demonstrate that our analysis is capable of re-producing results achieved by using explicit quadratic limb darkening equation, and that the filtering technique helps eliminating spurious spectral features in regions with strong telluric absorption. The method is applied to the VLT FORS2 observations of the exoplanets GJ 1214 b and WASP-49 b, and our results are in agreement with previous studies. Comparisons between obtained stellar specific intensity and numerical models indicates that the method is capable of accurately reconstructing the specific intensity. The proposed method enables more robust characterization of exoplanetary atmospheres by separating derivation of planetary transmission and stellar specific intensity spectra (that is model-independent) from chemical and physical interpretation.
\end{abstract}


\section{INTRODUCTION}
Transiting exoplanets offer a unique opportunity to study exoplanetary atmospheres. As a planet passes in front of its host star, a small fraction of the stellar light will travel through the semi-transparent atmosphere of the planet. Due to a wavelength-dependent optical thickness of the atmosphere, the fraction of stellar light able to penetrate the atmosphere will differ across the observed wavelength range. Measuring the change in stellar brightness during a transit event as a function of wavelength represents a way to detect and characterise exoplanetary atmospheres. Such measurements are often referred to as the planet transmission spectroscopy and described in terms of a wavelength-dependent planetary radius. Throughout a transit the planet blocks light originating from  different parts of the stellar disk. This enables indirect measurements of brightness variation on the stellar disk (limb darkening).

Even though the wavelength-dependent variations of effective planetary radius are small, on the order of a few percent at most, numerous successful measurements of exoplanet transmission spectra have been made. The most successful observational methods use narrowband filter photometry or low-resolution spectroscopy, which give high-precision transit light curves for a range of wavelength bins. A wavelength-dependent planetary radius is then found by fitting properties of the planet to the transit light curves. To account for stellar specific intensity variation across the stellar disk, the majority of the light-curve analysis methods rely on limb darkening equations \citep{Mandel2002ApJ, Gimenez2006A&A, Abubekerov2013MNRAS}. Limb darkening is then used to trace the stellar intensity along the path of the planet, thus the size of the planet can be obtained. However, limb darkening equations require coefficients, which are free parameters. The lack of spatial resolution leaves two options: either fitting these parameters together with properties of the planet, which can cause degeneracy between the size of planet and limb darkening coefficients, or deriving the limb darkening coefficients directly from stellar model atmospheres, which requires good knowledge of stellar parameters making results heavily model-dependent.

In this paper we explore an alternative approach for finding stellar specific intensity using a generalised formulation for limb darkening, which allows fitting observations in each spectral pixel without introducing degeneracy between planetary radius and specific intensity. The observations required for this method consist of a time series of flux-normalized spectra of a star during an exoplanet transit event. Observational techniques to achieve this are described in Section \ref{analysisMethod}. The analysis method itself is described in Section \ref{analysisMethod}. It consists of a two-stage fitting process of two unknowns: specific intensity (Section \ref{specificIntensity}) and exoplanet radius (Section \ref{radius}). Filtering of derived planetary radius with data-quality as weight is then preformed. Spectral resolution is sacrificed in order to homogenize the signal-to-noise ratio (SNR ) across the wavelength range (Section \ref{optimalFilter}). We test the method using synthetic observations, described in Section \ref{testing}. Requirements for observations and targets are discussed in Section \ref{requirements}. Finally, the method is applied to a set of FORS2 observations of the super-Earth GJ 1214 b and hot-Jupiter WASP-49 b, as described in Section \ref{gj1214b} and \ref{wasp49} respectively.

\section{DATA ANALYSIS METHOD} \label{analysisMethod}
\subsection{Overview} \label{analysisMethodOverview}
There are many parameters that contribute to the shape and depth of a transit light curve. Some of these parameters are wavelength-independent and can be determined from the white light (e.g. center of transit) or from previous radial velocity measurements of the same system (e.g. orbital parameters). We do not propose a new method for determining orbital parameters, nor does the proposed data analysis method require better knowledge of these than currently used methods. We therefore simply assume that all necessary orbital parameters (semi-major axis, orbital period, inclination, eccentricity, epoch of transit center) are known with "sufficient" precision. In Section \ref{planetPosition} we clarify the meaning of "sufficient" and show that our needs are compatible with routinely achieved precision. Planetary radius and stellar specific intensity will also affect the shape and depth of the light curve. Both are wavelength-dependent and thus not possible to determine beforehand or from white light. Herein lies the challenge: disentangling the wavelength-dependent specific intensity from planetary radius.

The data-set required for the proposed method consists of the wavelength dependent light curves of a star during a transit event. In this paper we will focus on ground based observations, using multi-object spectroscopy to monitor temporal changes in the telluric atmosphere. By observing an exoplanet hosting star, as well as several other non-variable stars simultaneously (preferably with similar spectral energy distribution as the exoplanet host star), temporal changes in telluric transmittance and instrument throughput can be monitored and accounted. This is done by normalizing the light curve of the exoplanet hosting star in each wavelength by the light curve of the (average) telluric standard star in the same wavelength. Any changes in the telluric standard stars (assuming they are not variable) should be a result of changes in the telluric atmosphere, and should have affected the target star in the same way.
	
Transit spectroscopy using low spectral resolution has an obvious advantage over high resolution; it is easier to reach the SNR such that the planetary radius can be measured with high enough precision to detect small variations due to the presence of an atmosphere. However, low spectral resolution makes the interpretation more dependent on a model of planetary atmospheres. Optimal results are obtained by finding the best trade-off between spectral resolution and precision. This is explored in Section \ref{optimalFilter}.
		
Once temporal changes not originating from the transiting exoplanet have been removed, each individual wavelength channel is normalized using the average flux in out-of-transit exposures at the same wavelength. This will result in a typical transit light curve for each spectral channel, with out-of-transit flux equal to 1 and a transit depth proportional to the fraction of stellar light blocked by the exoplanet. In Section \ref{syntObsSec} we show this reduction procedure implemented on synthetic data, and in Section \ref{GJ1214_DataReduction} we show this implemented on FORS2 observations.

In the first stage of the data analysis, the transit light curve in each wavelength channel is analyzed individually. From a first guess for planetary radius, the specific intensity (as function of disk position) that produces the best fit to the observed light curve is established. The guess for planetary radius is updated, and this will result in a different specific intensity that achieves a (different) best fit to the observed light curve. A smaller planet requires brighter parts of the star to be blocked to produce the same transit depth, and vice versa for a larger planet. The planetary radius at a given wavelength is then found by searching for the radius (and the corresponding specific intensity) that enable the best overall fit. 

Performing such fit for every wavelength bin results in a planetary transmission spectrum in form of effective planetary radius as a function of the wavelength. The SNR at some wavelengths will however be low, making the derived radius uncertain. Accuracy can be increased through sacrificing spectral resolution: in spectral regions with poor data quality spectral resolution is decreased compared to the regions with higher data quality. 

While central to our method this formulation for obtaining limb darkening can also be integrated in any of the existing transit light curve packages (e.g. TAP \citep{Gazak2011ascl}, PhoS-T \citep{Mislis2012hell}, EXOFAST \citep{Eastman2013PASP}, VARTOOLS \citep{Hartman2015IAUGA}, batman \citep{Kreidberg2015PASP}).


\subsection{Specific Intensity, Mathematical Definition} \label{specificIntensity}
There are many approaches to obtaining stellar specific intensity. Two of the most commonly used methods are limb darkening equations and spectral synthesis using theoretical model atmospheres. For most model grids the latter is known to have problems reproducing center to limb variations even for the Sun, and it is close to impossible to compare with other stars (see e.g. \citep{Czesla2015A&A} and the discussion therein). Limb darkening equations are effectively approximation formulas and as such they have free parameters. For the limb darkening law to work, the coefficients should be carefully selected for each wavelength bin. We expect them to be nearly constant in regions with low line absorption but they will change significantly across the profiles of strong lines.

In this paper we present an alternative approach for determining the specific intensity. Our method preserves the properties of specific intensity predicted by the stellar atmosphere theory and captured by limb darkening equations, but we do not use algebraic expression with uncertain free parameters. Instead we derive specific intensity as a function of the limb distance from the observed light curves using inverse problem approach. Our method handles each wavelength bin sufficiently independently to account for differences in the optical depth scales at different wavelengths.

We assume a spherically symmetric star, such that the intensity at any given point on the stellar disk can be uniquely described by the projected distance from the center of the star as seen by a distant observer. We assume that the spectral resolution of observations is low enough to ignore Doppler shifts induced by stellar rotation. In the following sections the wavelength dependence have been omitted from equations. All equations in this section are solved for each wavelength channel independently. 

Stellar disks appear brighter when radiation originating from deeper, and therefore hotter and brighter layers of the atmosphere is able to leave stellar surface. The optical path from a particular geometrical depth in the direction of an observer is inversely proportional to the cosine of the angle between the direction of propagation and the direction to the local zenith ($\mu$). In the center of the disk $\mu$ is 1 and the same optical depth corresponds to a deepest geometrical layer compared to all other points on stellar disk. Close to the limb, $\mu$ is small and radiation leaving the star will mostly originate from the upper cooler layers. For our purposes using the linear distance from the center of the star instead of $\mu$ is better. We call this quantity $r$ and measure it in stellar radii (i.e. $r$ is 0 at the center of the disk and 1 at the limb). 

We restrict possible solutions for $I$ by imposing four requirements that are quite natural, at least for the main sequence stars observed in the optical and infra-red wavelengths. We use these relations as constraints for the physical quantities (Equation \ref{prop1} and Equation \ref{prop4}) and their dependence on the limb distance (Equation \ref{prop2} and Equation \ref{prop3}). The additional constraints are combined with the minimization  problem that searches for the best fit to the observed monochromatic light curves. Some of the assumptions may not be valid if the transit path crosses a stellar spot or bright facular networks. We will address these cases in Section \ref{starspots}. Likewise in wavelengths associated with emission lines or other forms of limb brightening these assumptions no longer hold. In dwarf stars of solar type or cooler (the primary targets for transit spectroscopy observations), limb brightening will not occur in most of the optical to near infra-red. The imposed additional constraints on specific intensity are:

\begin{itemize}
	\item [1.] Intensity is always positive:
	\begin{equation} \label{prop1}
	I(r) > 0
	\end{equation}
	\item [2.] Brightness decreases towards the limb: 
	\begin{equation} \label{prop2}
	\frac{dI(r)}{dr} < 0
	\end{equation}
	\item [3.] The rate of brightness decrease increases towards the limb of the disk, due to the curvature of the surface of a spherical star:
	\begin{equation} \label{prop3}
	\frac{d^2I(r)}{dr^2} < 0
	\end{equation}
	\item [4.] Intensity integrated over the projected stellar disk equals the flux, assumed to be 1:
	\begin{equation} \label{prop4}
	2 \int \limits_{0}^{1} {I(r) r dr} = F = 1
	\end{equation}
\end{itemize}
Note that Equation \ref{prop2} and Equation \ref{prop3} reflect the properties of the limb darkening law. A conventional form for a quadratic limb approximation is $J(\mu) = 1 - \gamma_1 (1 - \mu) - \gamma_2 (1 - \mu)^2$, where $\mu$ is the cosine of the angle between the local zenith and the direction towards a distant observer, and relates to $r$ as $\mu = \sqrt{1-r^2}$. This form of quadratic limb darkening is selected because it ensures that the coefficients, $\gamma_1$ and $\gamma_2$, are both positive. It is easy to show that in this case first derivative with respect to $r$ will be negative for any coefficients and for all $r$. Likewise the second derivative will always be negative. I.e. all quadratic limb darkening functions are included in solutions we consider.

\subsection{Specific Intensity, Numerical Implementation} \label{specificIntensityNumerical}
We use the following procedure to ensure that Equation \ref{prop1}, \ref{prop2}, \ref{prop3} and \ref{prop4} are all satisfied. Specific intensity, $I(r_n)$, is reconstructed at n pre-defined equispaced distances from the center of the disk to the limb (note that these do not need to the locations of the planet during transit observations). Values of $I$ in the next point are defined using a step size from the previous value (closer to the center). Starting from some arbitrary intensity ($b$) at the center of the disk, we define sequential values of $I(r_n)$ as described by Equation \ref{defineI_1}. As long as the difference in step size ($s_n$) is positive, using a step size equal to the negative sum of all previous $s_n$ will satisfy both of Equation \ref{prop2} and Equation \ref{prop3}: Steps will always be negative and step size will always increase. The main advantage of using this formulation is to ensure that both Equation \ref{prop2} and Equation \ref{prop3} are satisfied automatically. The step size difference ($s$) has a unique value for each step (i.e. each $r_n$), and thus are a collection of free parameters, which together uniquely defines the intensity. 

The intensity outside the star ($r > 1$) is set to 0. Since the derivative is always negative, this forces the intensity function to be positive at all points (Equation \ref{prop1}). Note that Equation \ref{defineI_1} alone do not guarantee this. The function is then scaled to make the disk-integrated specific intensity match the stellar flux, see Equation \ref{defineI_2}. After this step, the normalized specific intensity ($i$) will satisfy all constraints for any positive values of $s_n$. This step ensures that the final result is insensitive to the initial guess for innermost point ($I(r_1) = b$).

\begin{align} \label{defineI_1}
& I(r_1) = b \nonumber \\
& I(r_2) = I(r_1) - s_{1} \nonumber \\
& I(r_3) = I(r_2) - (s_{1} + s_{2}) \nonumber \\ 
& ...\nonumber \\
& I(r_n) = I(r_{n-1}) - \sum^{n-1}_{k=1} s_k
\end{align}
\begin{equation} \label{defineI_2}
	i(r) = \frac{I(r) -I(r>1)}{2 \int \limits_{0}^{1} {(I(r)-I(r>1)) r dr}} 
\end{equation}

From the initial guess for starting points ($b$) and step size differences ($s_n$), a specific intensity function ($i(r)$) is computed. With this intensity, the first guess for the ratio of planetary to stellar radius ($R_p$) and previously established orbital parameters we calculate a synthetic transit light curve ($O_{synt}(\phi)$), using Equation \ref{syntObs}. The stellar flux ($F$) is set to 1 and specific intensity ($i(r)$) is integrated over the area of the stellar disk blocked by the planetary disk ($A_p(R_p, \phi)$). The transit light curve is sampled at the same orbital phases ($\phi$) as the observations ($O_{obs}(\phi)$), i.e. the same positions on the stellar disk. Once a synthetic light curve is created, the quality of the synthetic specific intensity can be assessed by calculating the differences squared ($\sigma^2$) between synthetic and observed light curves, see Equation \ref{residuals}.

\begin{equation} \label{syntObs}
O_{synt}(\phi) = F - \int\limits_{A_p(R_p,\phi)} i(r) dr 
\end{equation}

\begin{equation} \label{residuals}
\sigma ^2= \sum_{\phi} \big[ O_{obs}(\phi) - O_{synt}(\phi) \big]^2
\end{equation}

Changing $s_n$ will generate new {$i(r)$, $O_{synt}(\theta)$. Thus we can find the optimal set of $s_n$ that minimizes $\sigma^2$. The optimal $s_n$ corresponds to the specific intensity that produces the synthetic light curve with best fit (in least squares sense) to observed data for an assumed planetary radius. We choose to derive the initial guess for $s_n$ from fitting specific intensity to the white light curve (i.e. wavelength averaged), but initial guess could also be taken from limb darkening equations. See Appendix \ref{appendix} for details on numerical implementation. If SNR ratio differ between flux measurements, one could introduce a weight in Equation \ref{residuals} that affects the relative importance of each exposure, such that exposures with low SNR contribute less to $\sigma$ than exposures with high SNR.

\subsection{Planetary Radius}  \label{radius}
Changing the assumed planetary radius ($r_p$) in Section~\ref{specificIntensity} will result in a different solution for specific intensity. A larger planet requires a fainter part of the star to be blocked in order to produce the same observed transit depth, and a smaller planet requires a brighter part of the star to be blocked. At first glance this appears to be a degenerate problem, where changes in planetary radius can be compensated by the specific intensity. However, the defined properties of specific intensity in Section \ref{specificIntensity}, most importantly Equation \ref{prop4} in this case, limit the range of possible specific intensity functions, ensuring a unique solution for each given planetary radius. This is demonstrated in Figure~\ref{FitIntensty}, where we search for the best fitting specific intensity to an artificial transit light curve (created using the method outlined in Section~\ref{syntObsSec}), using three different assumed planetary radii. As shown in the figure, not only the transit depth, but also the shape of the light curve changes with the planetary radius. This ensures a unique solution for each given planetary radius, breaking the apparent degeneracy. Note that if the data quality is poor or if planetary radius or specific intensity are not sampled sufficiently, one cannot eliminate the possibility of local minima.

The quality of the fitted parameters are quantified by $\sigma^2$, which has a unique value for each assumed planetary radius. The planetary radius with best fit to observations is found by searching for the radius (and corresponding specific intensity and synthetic light curve) that minimizes $\sigma^2$.

\begin{figure}[h]
	\centering
	\includegraphics[width=\hsize]{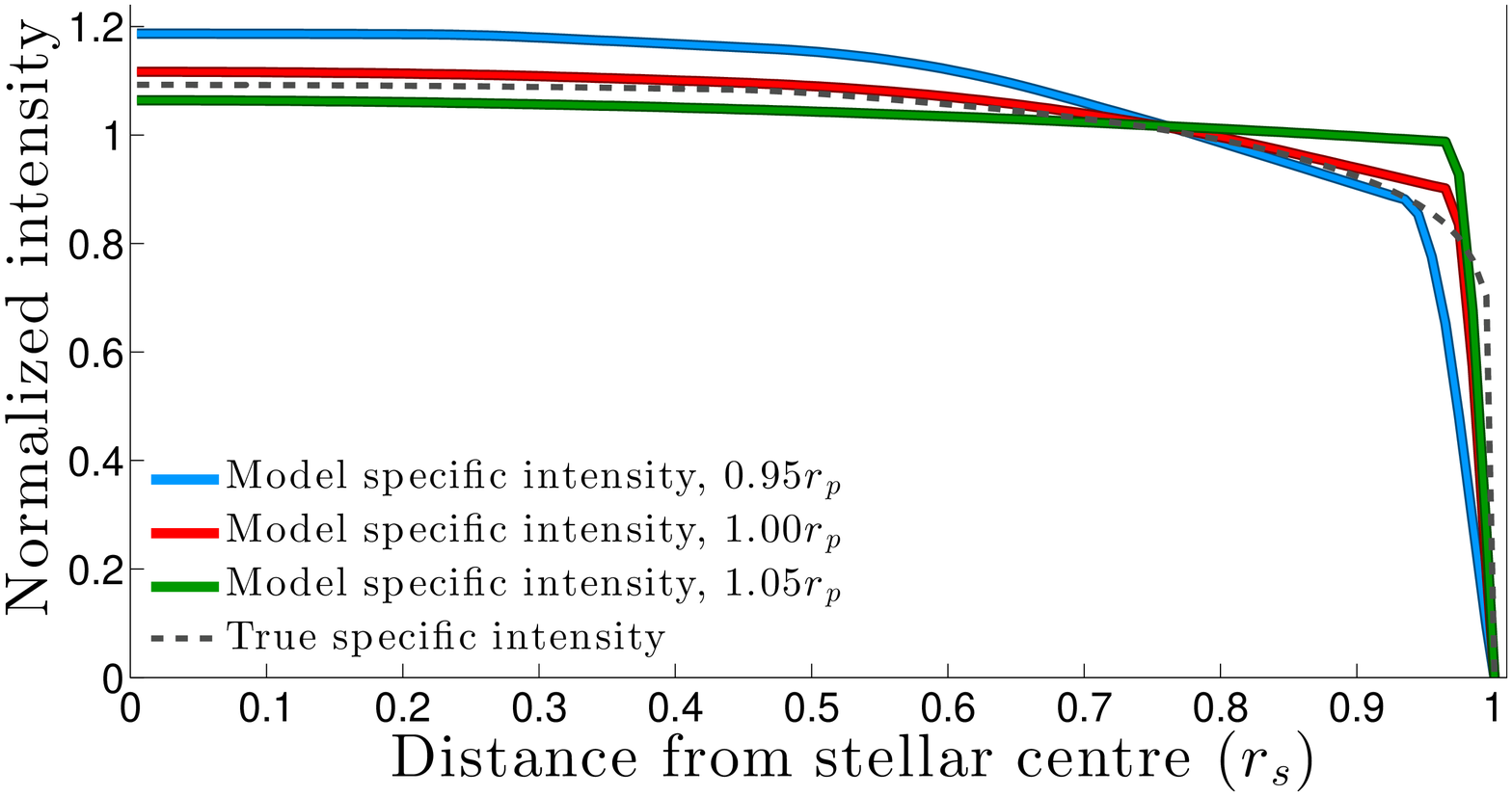}
	\includegraphics[width=\hsize]{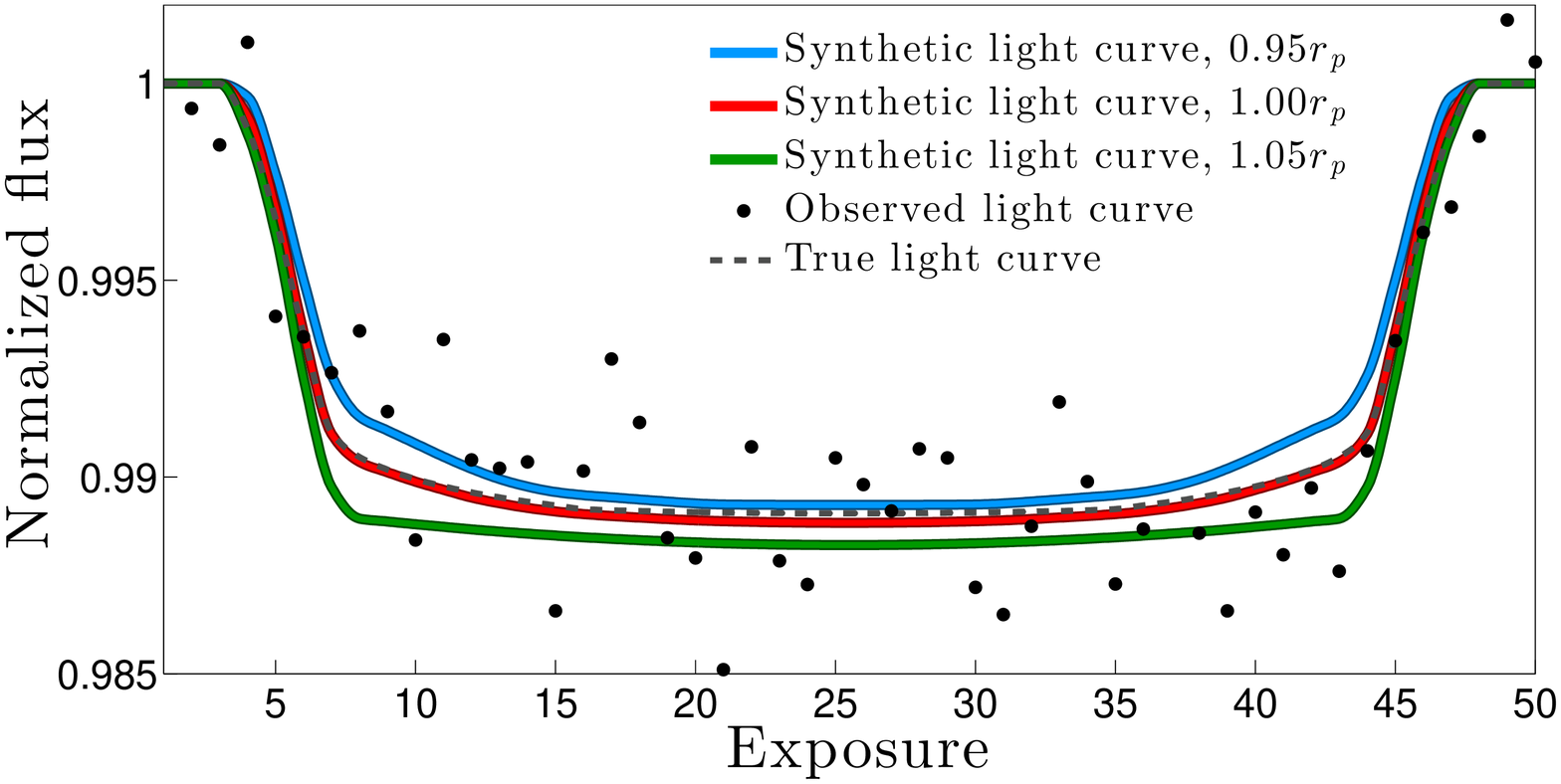}
	\includegraphics[width=\hsize]{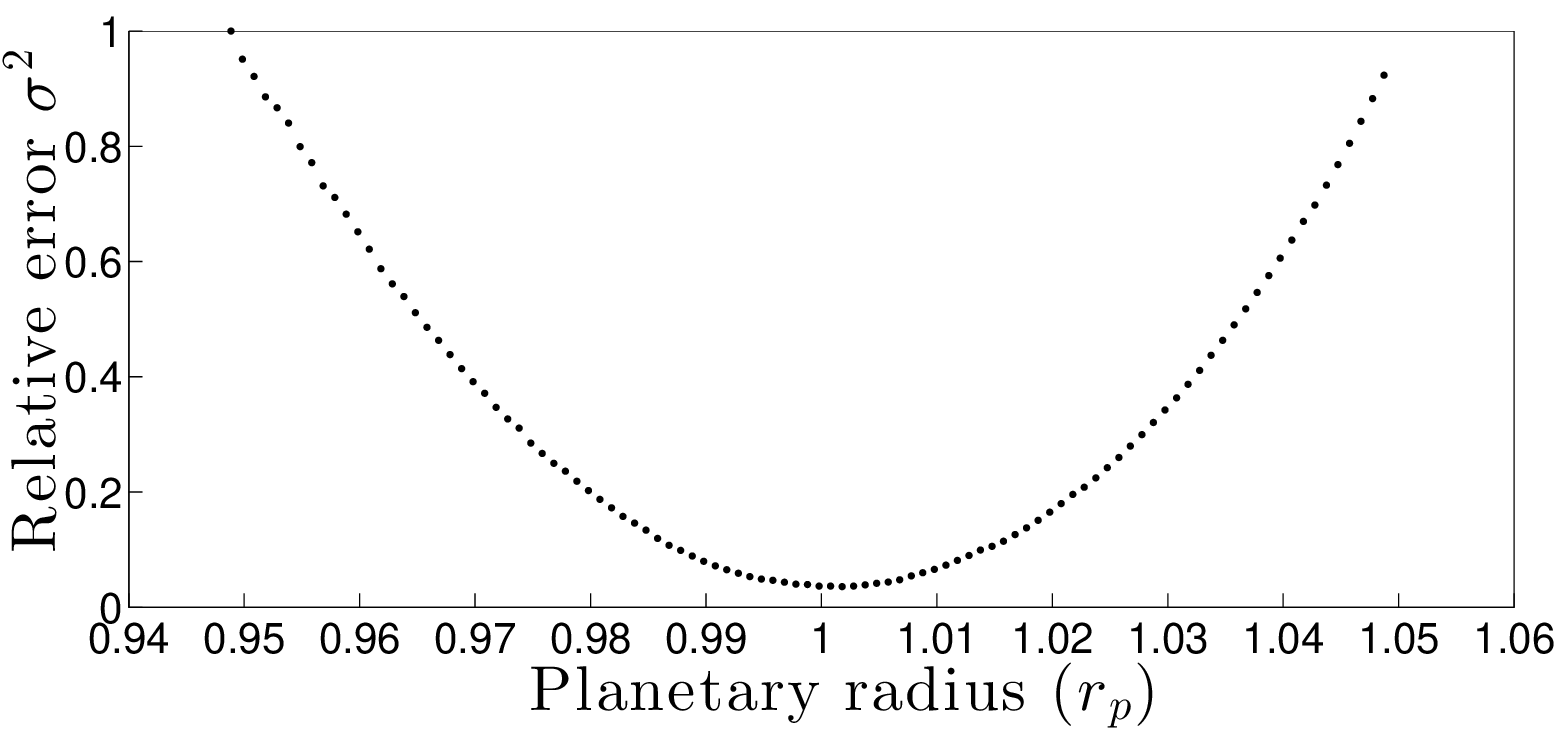}
	\caption{Upper panel: Best fitting specific intensity model for three different assumed planetary radii. The specific intensity used for creating artificial observation is shown as a dashed line. Stellar model taken from case 2 in Table~\ref{planetPrameters} at $\lambda = 1.60 \mu m$. Note that the x-axis is projected linear distance from the center of the star, rather than the more commonly used $\mu$-angle. \newline 
	Middle panel: Synthetic light curves from derived specific intensity and assumed planetary radius. Artificial observation is shown as black dots, and theoretical noiseless transit curve is shown as a dashed line.\newline
	Lower panel: Differences squared between observed flux and best possible fit increase as assumed planetary radius moves away from $R_p=1$. The best possible fit is achieved close to the correct value.}
	\label{FitIntensty}
\end{figure}

\subsection{Wavelength dependence}  \label{optimalFilter}
Performing the two steps described in Section~\ref{specificIntensity} and \ref{radius} for each wavelength channel individually will generate the planetary radius as a function of wavelength, $r_{p}(\lambda)$. However, reaching the same fidelity level for the planetary radius in each individual wavelength channel (even at spectral resolution $\lambda / \Delta \lambda \approx 500$) may be impossible. Given that exoplanet transits are time-limited events, increasing integration time is not a viable option. Instead combining data from neighboring wavelength points is often used to increase the accuracy of derived results. Data quality (and therefore accuracy of derived results) will vary across the spectral range. For example, for ground-based observations telluric absorption will determine the quality of radius reconstruction. By allowing the spectral resolution to vary across the observed wavelength range, the discrepancies in quality of the derived radius can be reduced. We achieve this by using an optimal filtering technique with Tikhonov first order regularization \citep{TikhonovArsenin77}, Equation \ref{Tikho1}. 
first order regularization \citep{TikhonovArsenin77}, Equation \ref{Tikho1}. 

The solution to Equation \ref{Tikho1} is found by searching for the function $R_p$ (corresponding to planetary radius) that minimizes $\Phi$. The first term in the equation strives to keep function ($R_p$) close to the derived planetary radius $r_p$. The second regularization term forces $R_p$ to be smooth by forcing the differences between adjacent wavelength bins to remain small. We use the inverse of the squared residuals ($1 / \sigma(\lambda)^2$) as a weight. At wavelengths where $1 / \sigma^2$ is small (corresponding to derived planetary radius with poor fit to observed data), the regularization terms will dominate, forcing a smooth solution. At wavelengths where $1 / \sigma^2$ is large, regularization is less important and the observational data will determine the solution. The relative importance of regularization is controlled by a free parameter, $\alpha$. This regularization parameter can be tuned to match the overall data quality (small values for high SNR and large values for low SNR). $R_p$ is found by setting derivatives of $\Phi$ to zero and linearizing the derivatives in the regularization term. This creates a linear system of equations which is straightforward to solve. 

\begin{align} \label{Tikho1}
	\Phi = \sum_{\lambda} \Bigg[ \frac{1}{\sigma(\lambda)^2} \Big( R_{p}(\lambda) - r_{p}(\lambda) \Big) \Bigg]^2 + \\ \nonumber
	\alpha \sum_{\lambda} \Bigg[ \frac{d R_p(\lambda)}{d \lambda} \Bigg] ^2 = min
\end{align} 

The effect of the proposed regularization is to trade spectral resolution for reduction of extensive noise-induced oscillations. This is similar to the methods commonly used in the field: rebinning the data and reducing the resolution by a factor of 10 to 100. Both methods combine data from neighboring points, smoothing the spectrum and suppressing sharp and narrow spectral features. The advantage of our method is that it offers a controlled balance between spectral resolution and accuracy.

\subsection{Finding Optimal Regularization Parameter}  \label{optimalregPara}
Finding the optimal regularization parameter ($\alpha$) is not a trivial task. Small values of $\alpha$ result in better fit to observations but the recovered solution for the radius will have large and unphysical  oscillations around the true value. With increasing value of $\alpha$ the recovered solution approaches the mean value but eventually the real spectral features will be lost. Finding the regularization that achieves the optimal balance between recovering the actual spectroscopic features in the solution without fitting the observational noise a priori is often impossible. In this section we describe a method that can help determining the optimal value of $\alpha$ for our particular problem. In order to assess the trade-off between matching the observations and the smoothness of the solution we need a way to compare the two.

(1) For estimating the quality-of-fit, we construct the residuals of observation minus synthetic observations, where both the observations and the synthesis are subject to optimal filtering (i.e. both are smoothened in the same way). The synthesis is constructed according to Equation~\ref{syntObs}, using the previously recovered specific intensity ($i(r)$) and unregularized planetary radius ($r_p$). Obviously the fit will improve with the increase of $\alpha$.
	
(2) For estimating the reliability of the recovered solution we again compute the residuals of observations minus synthesis, except now neither observations nor synthesis are regularized. Instead the synthesis is created (Equation~\ref{syntObs}) using the regularized planetary radius ($R_p$). This serves as a test for how important sharp details in the planetary radius spectrum are for the reproduction of observations. In this case the residuals will increase with increasing $\alpha$.

In case (1), both observations and synthesis are filtered following Equation \ref{Tikho1} except we replace $r_p(\lambda)$ with the observed or synthetic flux $e(\lambda)$ obtained in a single exposure. The target is the regularized flux $E_{reg}(\lambda)$ that replaces the regularized planetary radius ($R_p(\lambda)$). This results in Equation \ref{RegObservation}, where the weight ($1 / \sigma(\lambda)^2$) is the same weights as used in Equation \ref{Tikho1}. We obtain the regularized observation by searching for the $E_{reg}(\lambda)$ that minimizes $\Theta$.

\begin{align} \label{RegObservation}
	\Theta = \sum_{\lambda} \Bigg[ \frac{1}{\sigma(\lambda)^2} \Big( E_{reg}(\lambda) - e(\lambda) \Big) \Bigg]^2 + \\ \nonumber
	\alpha \sum_{\lambda} \Bigg[ \frac{d E_{reg}(\lambda)}{d \lambda} \Bigg] ^2 = min
\end{align} 

This process gives us a way to assess how the quality-of-fit (1) and the reliability of the recovered solution (2) are affected by the choice of regularization parameter. The (close to) optimal regularization parameter is the $\alpha$ that yields the minimum when summing squared residuals for (1) and (2). We use this value to regularize the planetary radius to construct the final solution. Further exploration of this method is done in Section~\ref{detemineRegPara}.

\section{APPLICATION - SIMULATED DATA}
In this section we test the data analysis method by applying it to synthetic transit observations. Use of synthetic data enables reliable assessment of limitations and requirements to observations as well as to the analysis method, since the true values for all fitted parameters are known.

\subsection{Data Sources} \label{dataSources}
\subsubsection{Exoplanet Spectra}
We study the method using medium/low-resolution transit spectroscopy from the surface of the Earth. This makes water very difficult to detect due to strong absorption from water vapour in the telluric atmosphere. We use a Venus-like exoatmosphere, for which the transmission spectrum is dominated by CO$_2$ absorption. The size of the exoplanet ($R_e$) was set to 2.5 $R_\oplus$, a typical size for a super-Earth. The atmospheric transmittance was computed using LinePak \citep{Gordley1994} along the rays crossing the atmosphere at increasing distances from planetary center. Integration over projected area of corresponding concentric rings gives the total transmittance of planetary atmosphere.

\subsubsection{Stellar Spectra}
Synthetic spectra generated with the stellar atmosphere modeling code MARCS \citep{Gustafsson2008} were used as "observations" for both the star in the transiting system and for telluric standard stars. For the latter, only the flux spectrum was required, while for the exoplanet host star, both flux and specific intensity were computed.

To highlight the importance of the size ratio between the exoplanet and its host star, we test the method using four host stars of different sizes, but all with the same super-Earth exoplanet, see Table~\ref{planetPrameters}. The planet was put at a distance from the star where it would receive the same irradiance as Venus does form the Sun. Orbital period and transit duration were then calculated based on this distance, assuming circular orbit and $90^{\circ}$ inclination. 
\begin{table}[h]
	\caption{Planet and stellar parameters for simulated observations.}
	\centering
	\begin{tabular}{lllll}
		\hline \hline
		& Case 1 & Case 2 & Case 3 & Case 4 \\
		\hline 
		$\mathrm{R_{s} / R_{\odot}} $ & 0.2 & 0.3 & 0.5 & 0.75 \\
		$\mathrm{R_{e} / R_{\oplus}} $ & 2.5 & 2.5 & 2.5 & 2.5 \\
		$\mathrm{R_{e}^2 / R_{s}^2} $ & 0.013 & 0.0058 & 0.0021 & 0.00093\\
		K mag & 12.0 & 12.0 & 12.0 & 12.0\\ 
		Semi-major & & & &\\
		Axis (au) & 0.038 & 0.078 & 0.16 & 0.38 \\
		Transit dur- & & & \\
		ation (min) & 68 & 119 & 220 & 400 \\
		Effective  & & & &\\
		temperature (K) & 3100 & 3300 & 3600 & 5000 \\
		Surface gra- & & & &\\
		vity (log g (cgs)) & 5.0 & 5.0 & 5.0 & 4.5 \\
		Metallicity & 0.0 & 0.0 & 0.0 & 0.0 \\ 
		Mass ($\mathrm{M_{\odot}}$) & 0.15 & 0.3 & 0.5 & 0.8 \\
		\hline
	\end{tabular}
	\label{planetPrameters}
\end{table}

\subsubsection{Telluric Transmission} \label{tellTrans}
Telluric transmittance was calculated using LinePak \citep{Gordley1994}, with an altitude set to 2.6 km (which is the height of Paranal observatory). Two types of temporal variations in telluric transmission were implemented, changes in precipitable water vapor (PWV) and target airmass. PWV starts at 2.5 mm (which is the median value at Cerro Paranal) and increases linearly until PWV has doubled at the end of the observation run. Airmass starts at 1.05 and increases to 1.25.

\subsection{Synthesizing Observations} \label{syntObsSec}
Observations were synthesized for a theoretical multi-object spectrograph at an 8m class telescope. The instrument was assumed to cover most of H and K bands ($1.5 - 2.4 \mu m$) in a single exposure, with a spectral resolution $\lambda / \Delta \lambda = 1800$. Integration time was kept constant during the transit and was selected to accumulate a total of 50 exposures during the transit and 50 exposures before and after the transit (translating to between 80 and 480 s/exp for the shortest and longest transit duration respectively). Wavelength dependent SNR was calculated using the  K-band Multi Object Spectrograph (KMOS ) \citep{Sharples2013Msngr}	Exposure Time Calculator (ETC) (www.eso.org/observing/etc), which is a proxy instrument for our simulations in terms of spectral coverage, photon collecting area and spectral resolution. We also synthesize observations of five telluric standard stars (between 4000 to 5500 K, and K mag 12 to 14). All observations are assumed to be affected be the telluric atmosphere in the same way, with temporal variations as described in Section~\ref{tellTrans}. 

We then process the observations as described in Section~\ref{analysisMethodOverview}. Critical stages of the analysis are illustrated in Figure~\ref{obsReduction}. At this scaling all visible lines have telluric or stellar origin. Temporal changes in telluric atmosphere have been exaggerated to make the changes easily visible to the reader. The reduced observations show a clear transit light curve pattern for most wavelengths. Beyond 2.4 $\mu m$ strong telluric absorption makes the data unusable, this region was removed from the data-set before further analysis was preformed. 

\begin{figure}[h]
	\centering
	\includegraphics[width=\hsize]{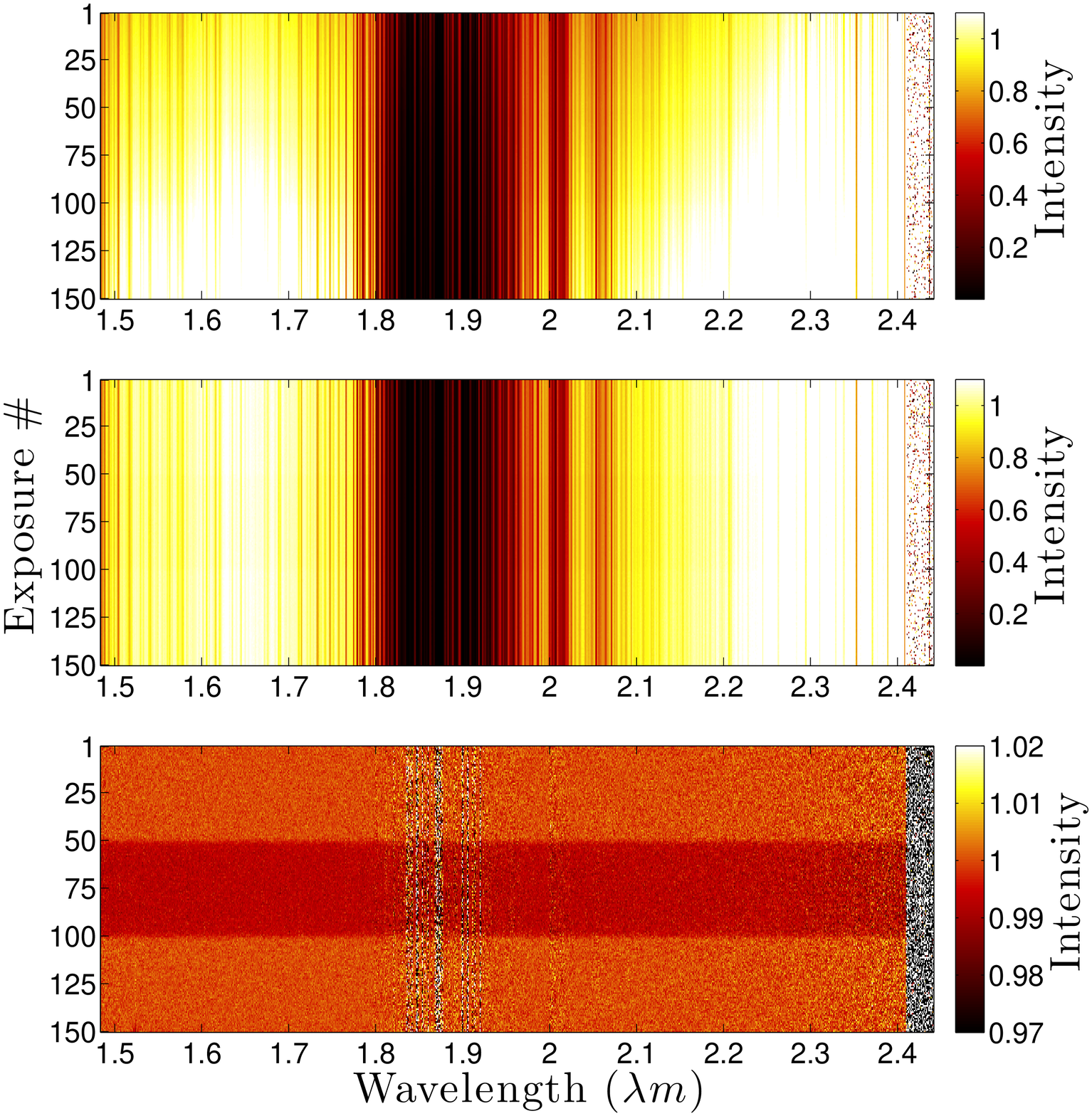}
	\caption{Data reduction procedure shown for synthetic observation of super-Earth transiting M-dwarf. \newline
		Upper panel: Observations after initial reduction (flat fielding, bias removal etc). \newline
		Middle panel: Removal of temporal changes, derived from telluric standard stars. \newline
		Lower panel: Normalization by average flux in out-of-transit exposures.}
	\label{obsReduction}
\end{figure}

\subsection{Recovery of Exoplanet Transmittance} \label{testing}
The data analysis method described in Section \ref{analysisMethod} was applied to synthetic observations of four transiting super-Earths (see Table. \ref{planetPrameters}). Results are shown in Figure \ref{results1}. We quantify the quality of the recovered planetary radius through $q$, which is defined as the sum of the squared differences between the recovered solution and the input model, normalized by this value for case 2 (where $R_s = 0.3 R_{\odot}$), as this is used as a reference case in Section~\ref{requirements}. For the two smaller host stars (0.2 and 0.3 $R_{\odot}$), the major absorption bands around 1.6$\mu m$ and 2.0$\mu m$ (both from CO$_2$) are clearly visible in reconstructed transmission spectra. For the two larger host stars (0.5 and 0.75 $R_{\odot}$) results are significantly worse. The average size of the planet was correctly recovered, but all spectral features were lost. This test demonstrates the well-known sensitivity of transit spectroscopy to the planet-to-star size ratio. We convert the derived planetary radius (in units of stellar radii) to km in order to make comparisons between host stars of different sizes. The data analysis method only derives the relative size of planet so the conversion to absolute units requires knowledge of the size of star, which can be obtained by for example using accurate distances and asteroseismology. Note that the size ratio of Jupiter to the Sun is roughly the same as the size ratio between the super-Earth and the smallest star used here, making transit spectroscopy of hot-Jupiters transiting Solar like stars feasible with this technique.

\begin{figure}
	\centering
	\includegraphics[width=\hsize]{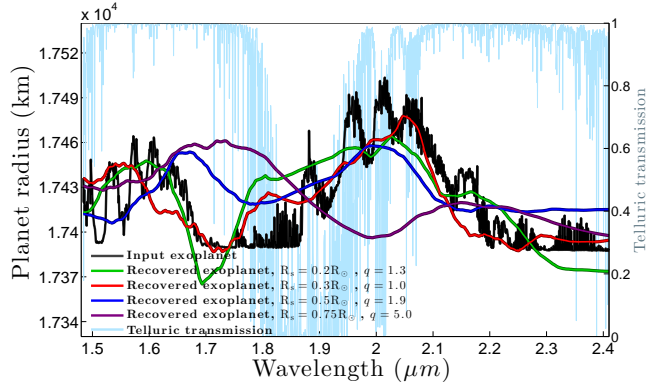}
	\caption{Tests of data analysis method with synthetic observations of four different systems. All with the same super-Earth but different sizes of host stars (see Table~\ref{planetPrameters}). For the two smaller stars (green and red line) the major spectral features are recovered. For the two larger stars (blue and purple line) real spectral features can no longer be distinguished from the spurious features.}
	\label{results1}
\end{figure}

\section{REQUIREMENTS} \label{requirements}
In this section we test and evaluate the effects imperfections in observations and underlaying assumptions has on the recovered solution. All tests in this section are performed for case 2 in Table~\ref{planetPrameters} (i.e. $R_s = 0.3 R_\odot$). In each test a single aspect of the observations is altered, and the recovered solution is compared with the reconstruction using ideal data.

\subsection{Orbital Inclination} \label{OrbitalInclination} 
If a transiting planet has an orbital inclination different than 90$^\circ$, the planet will not cross the center of the stellar disk, thus there is no information about specific intensity in this region. This lessens the constraints on specific intensity during the fitting procedure, making derived specific intensity less certain. However, we known that specific intensity changes slowly in central parts of the disk, i.e. for a given wavelength $I(r)$ is quite flat in the central regions. Thus orbits with inclinations (relative to the line of sight) slightly different from 90$^\circ$ should not compromise the derived specific intensity. Planets that only transits the outer parts of the stellar disk will be problematic. Results from the tests with inclination of 90.0$^\circ$, 89.9$^\circ$ and 89.5$^\circ$ are shown in Figure~\ref{test_incl}, along with visual representation of transit geometry for each case. Note that limb darkening equations suffer from the same problem, where lack of information of the intensity in large fractions of the stellar disk induces errors or requires limb darkening coefficients to be derived from stellar models.

\begin{figure}
	\centering
	\includegraphics[width=\hsize]{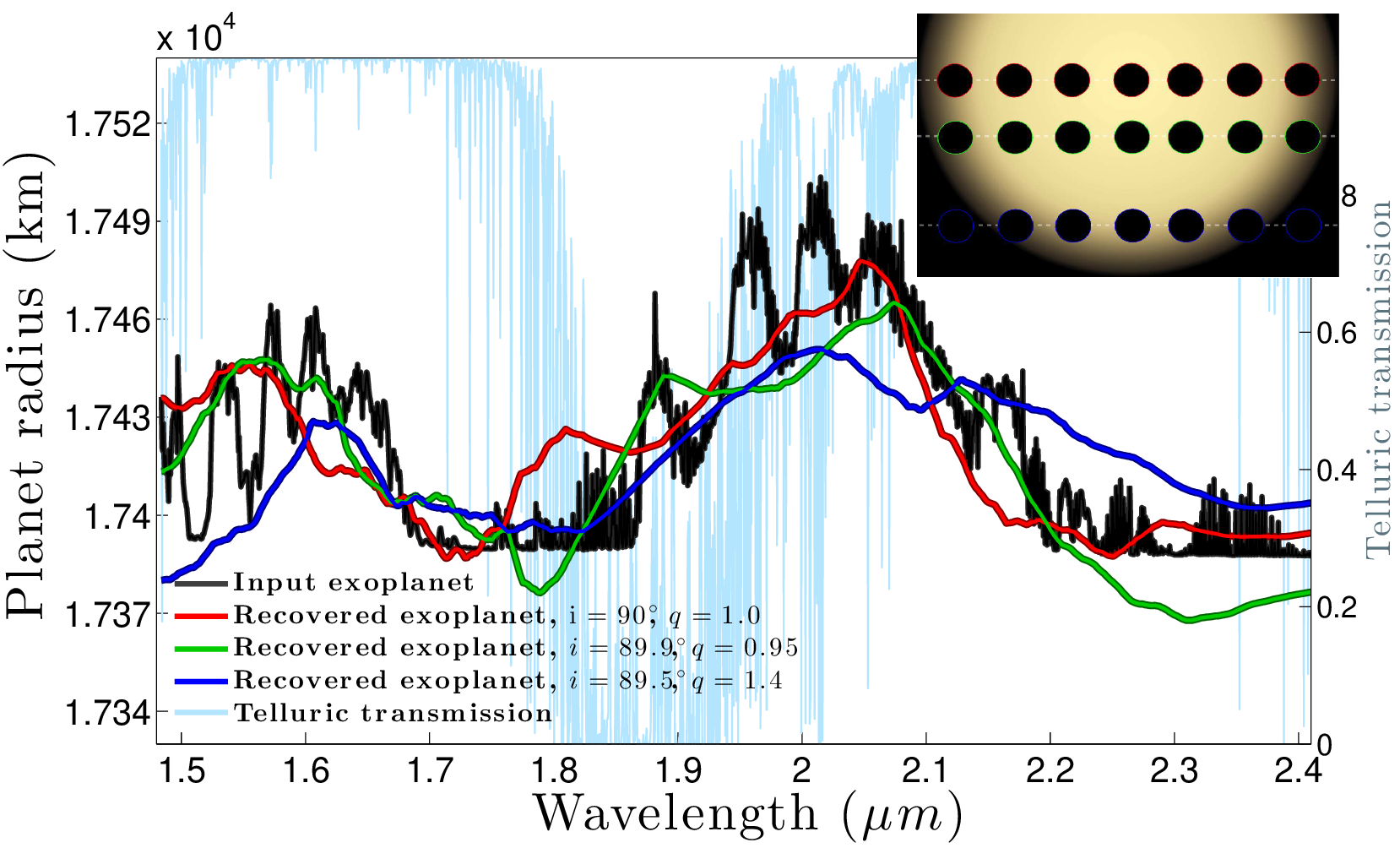}
	\caption{Tests of the effects of orbital inclination on the recovered transmission spectrum. \newline
		In red, reference case, inclination $90.0^\circ$. \newline
		In green, inclination $89.9^\circ$. \newline
		In blue, inclination $89.5^\circ$. \newline
		Inserted panel: Visual representation of the transiting planet's path across the stellar disk, from top to bottom, 90.0$^\circ$, 89.9$^\circ$ and 89.5$^\circ$ .}
	\label{test_incl}
\end{figure}

\subsection{Star Spots} \label{starspots}
We have postulated four requirements to constrain the specific intensity (Equation \ref{prop1}, \ref{prop2}, \ref{prop3} and \ref{prop4}). If a planet crosses a star spot during a transit, requirements 2-4 may no longer be fulfilled. In the presence of spots, the intensity would not continuously increase from limb to disk center; it may instead have a local minimum inside the spot. In transit light curves this can be observed as an increase of the flux as the planet crosses a spot. This effect has been detected in many transit light curves \citep{Pont2007A&A, Carter2011ApJ, Sing2011MNRAS, Nascimbeni2015A&A}. Ideally all exposures showing evidence of spot crossings should be removed from the data-set. In this case the derived results are not seriously degraded except for the reduced SNR due to fewer exposures, and a minor shift towards an overall larger planet. If spots are not detected (e.g the brightness change is below the noise level), contaminated exposures cannot be removed. However, this also means that the departures from the spot-free star are small (assuming SNR is high enough to detect planetary atmosphere). In this case the size of the planet may be slightly underestimated, but the wavelength dependence could still be recovered. If the star has a significant fraction of its surface covered in spots, but no spots are crossed by the planet, the specific intensity under the planet (with no spots) will be different from the average specific intensity contributing to the stellar flux. In such cases the size of the planet will be overestimated, but wavelength variations in the radius would still be possible to recover. All the above described cases were tested by applying the analysis method to synthetic observations. In the first case, 10\% of the exposures were contaminated by spots being crossed by the planet. These exposures were removed before applying the data analysis method . In the second case, 10\% of the exposures were contaminated by spots , but no exposures were removed. Spots were set to be of sizes comparable to the planet and 1000 K cooler than the surrounding photosphere. In the third case, no spots were crossed, but uncrossed regions of the star were covered in spots, enough to decrease the average stellar intensity by 1\% in regions that were not crossed by the planet (compared to regions that were crossed by the planet). Results shown in Figure~\ref{test_spots}.

Bright facular networks can have a similar effects. Just as spots, facular networks alter the brightness variations on the stellar surface such that the method outlined in Section \ref{analysisMethod} is not able to fully describe the parts of the star crossed by the transiting planet. As shown by \citep{Oshagh2014AA}, facular networks can lead to misinterpreting an observed increase in planet to stellar brightness ratio with shorter wavelengths as Rayleigh scattering the planetary atmosphere. While facular networks can cover a larger fraction of the stellar surface than stellar spots, the brightness contrasts is typically smaller. The method proposed in this paper should not be affected by potential errors from facular networks more than current standard methods. We therefore do not expect any major problems resulting from facular networks.

\begin{figure} [h]
	\centering
	\includegraphics[width=\hsize]{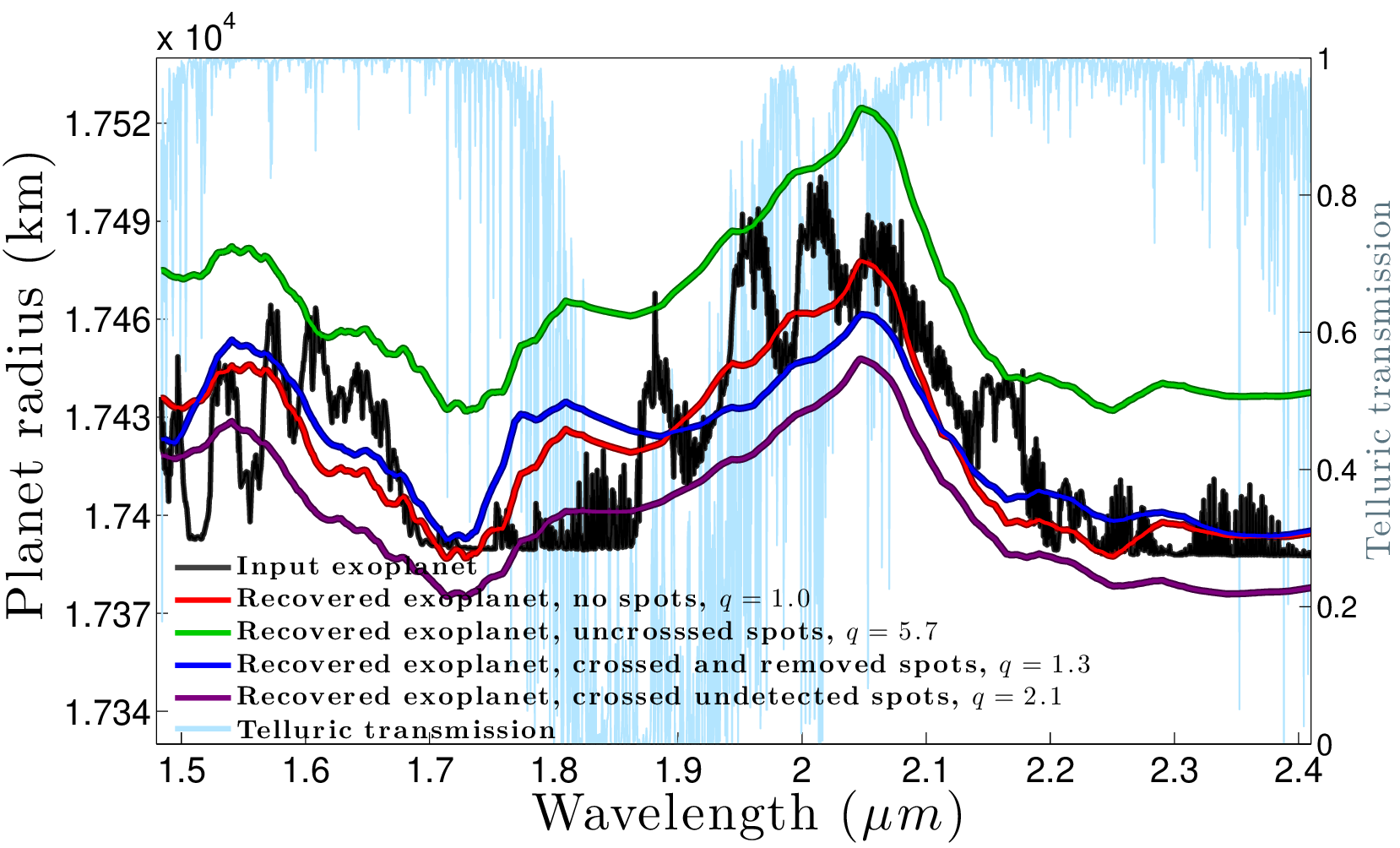}
	\caption{Tests of the effects of star spots on the recovered transmission spectrum. \newline
		In red, reference case, spot free star. \newline
		In green, the star is covered in spots, but the planet does not cross any of them. Difference between average intensity in the region where the planet transits and the overall intensity is on the order of 0.5\%. \newline
		In blue, the planet crosses spots, which were identified in the transit light curve and removed from the data-set before proceeding with the data analysis.\newline
		In purple, the planet crosses smaller spots, which could not be identified in the transit light curve and thus all exposures was included in the data analysis.}
	\label{test_spots}
\end{figure}

\subsection{Position of Exoplanet} \label{planetPosition}
The fitting procedure for stellar specific intensity relies on knowing the position of the transiting planet (its distance from the center of the stellar disk) at each given exposure. The effects of errors in assumed position of the planet is tested here in two ways. First by introducing a 0.1$^{\circ}$ error in orbital inclination, from 90.0$^\circ$ to 89.9$^\circ$, giving an offset in the assumed position (see inserted panel in Fig \ref{test_incl}). And secondly by offsetting the center of the transit by 90 seconds, shifting the planet's assumed position slightly. Results from these tests are shown in Figure~\ref{test_dist} and indicate that minor errors in assumed position of the planet will not affect the results significantly. The same spectral features can be recovered in all three cases.

\begin{figure} [h]
	\centering
	\includegraphics[width=\hsize]{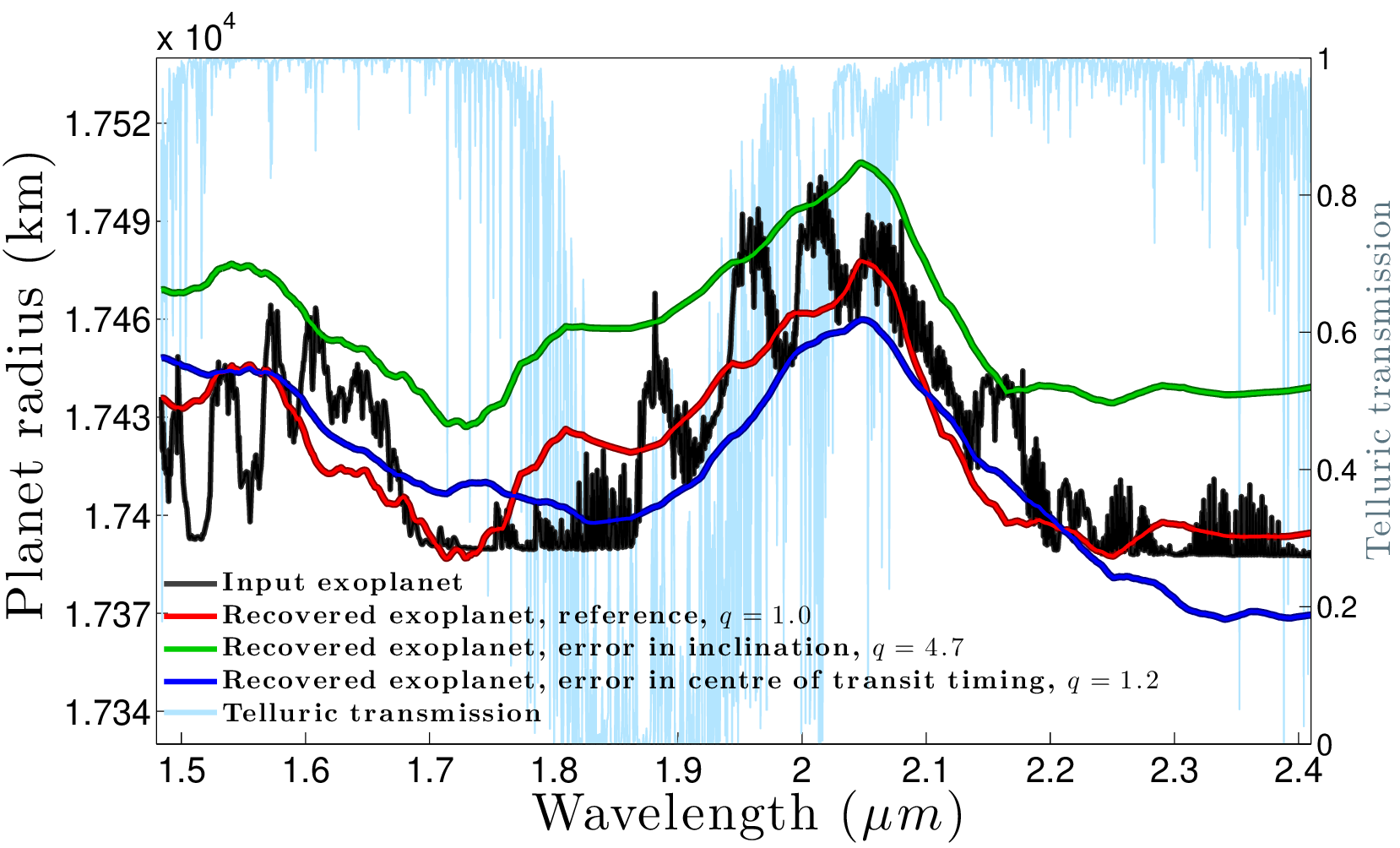}
	\caption{Tests of the effects of errors in the assumed position of the exoplanet on the recovered transmission spectrum. \newline
		In red, reference case, no errors. \newline
		In green, the orbital inclination used to derive the planet's position is changed with -0.1$^{\circ}$, resulting offset of assumed position. \newline
		In blue, the center of transit is assumed to be 90 seconds later than when simulating the observations.}
	\label{test_dist}
\end{figure}

\subsection{Comparison with Quadratic Limb Darkening}
We compare the proposed data analysis method with a more classical approach to transit spectroscopy. The method described in Section~\ref{analysisMethod} is compared with two slightly different versions based on quadratic limb darkening equations to obtain the specific intensity. In the first case, we derive the planetary radius in each spectral bin separately and then apply the optimal filtering technique as described in Section~\ref{optimalFilter}. In the second case we bin each 50 neighboring spectral pixels into a single data-point, and derive planetary radius using this data. In both cases the limb darkening is sampled over 20 wavelength points across the observed wavelength region, and coefficients for limb darkening equations are taken directly from the stellar model used to synthesize the observations (i.e. coefficients are optimal, but limb darkening is at lower spectral resolution than the observations). Results are shown in Figure~\ref{limbEqua}. This shows that the proposed method for obtaining specific intensity gives results comparable with limb darkening equations (when using optimal coefficients), and that the optimal filtering technique is a very useful tool for removing spurious features originating from regions with strong telluric absorption. It should however be noted that in terms of computational speed, quadratic limb darkening is significantly faster.

To demonstrate that our method for obtaining stellar specific intensity is capable of recreating stellar center-to-limb intensity variations with good accuracy without a need for functional form and free parameters for limb darkening equation, we tested both methods on identical synthetic observations of a transit event, in a single wavelength with high signal-to-noise ratio. Results are show in Figure~\ref{specIntVStest}. Our method better recreates the model, which is not surprising since quadratic limb darkening is a subset of our method (see Section~\ref{specificIntensity}) and it does not perfectly describe the input model. 

\begin{figure} [h]
	\centering
	\includegraphics[width=\hsize]{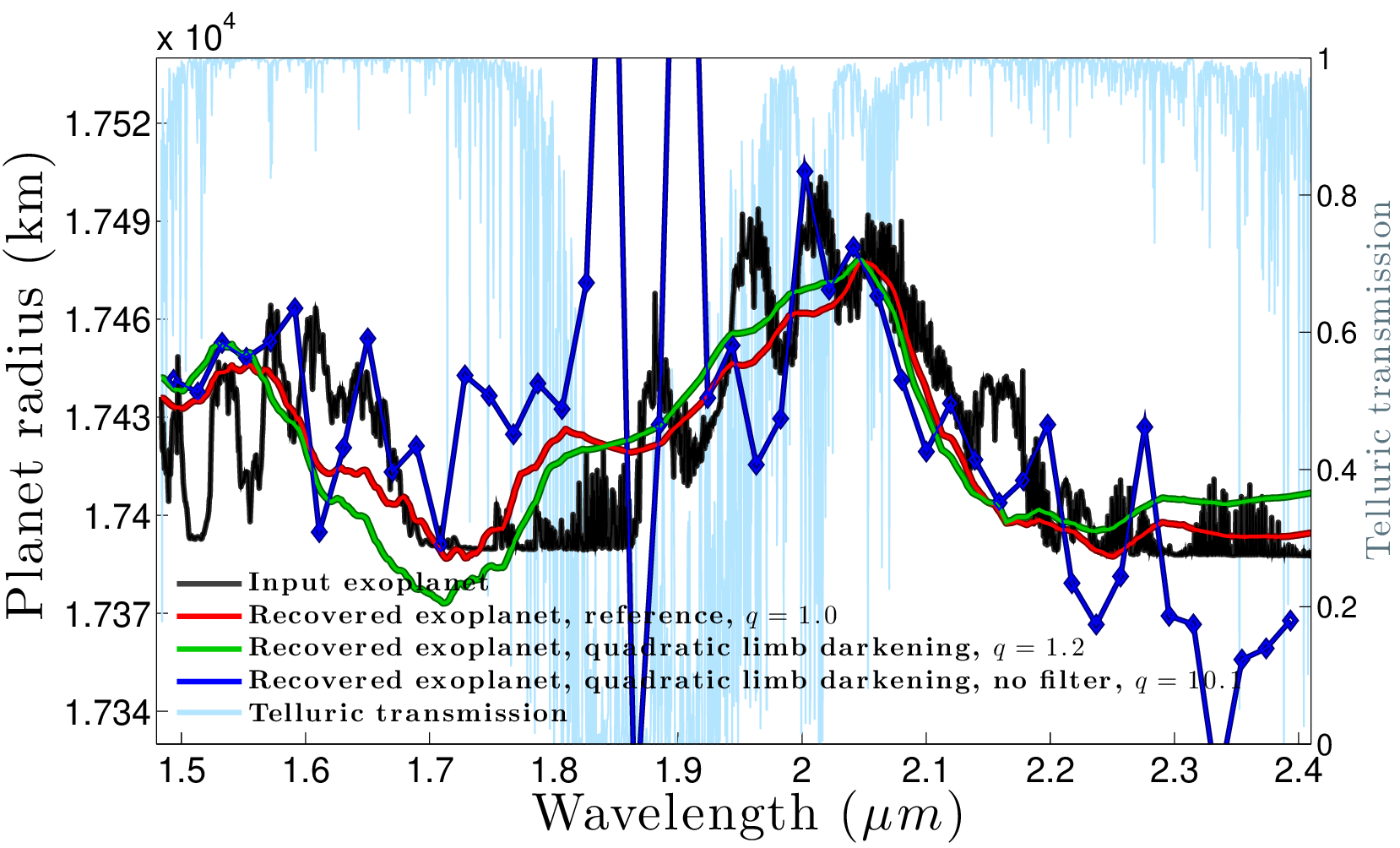}
	\caption{Tests of limb darkening equations (with optimal coefficients) compared with the method described in this paper.\newline
		In red, the inverse method approach described in this paper, including optimal filtering.\newline
		In green, limb darkening equation, with optimal filtering. \newline
		In blue, limb darkening equation, no optimal filtering, instead the data is binned over 50 neighboring points. The large jumps around 1.9 $\mu$m hs its origin in telluric contamination. It is not seen in the other cases due to optimal filtering.}
	\label{limbEqua}
\end{figure}

\begin{figure} [h]
	\centering
	\includegraphics[width=\hsize]{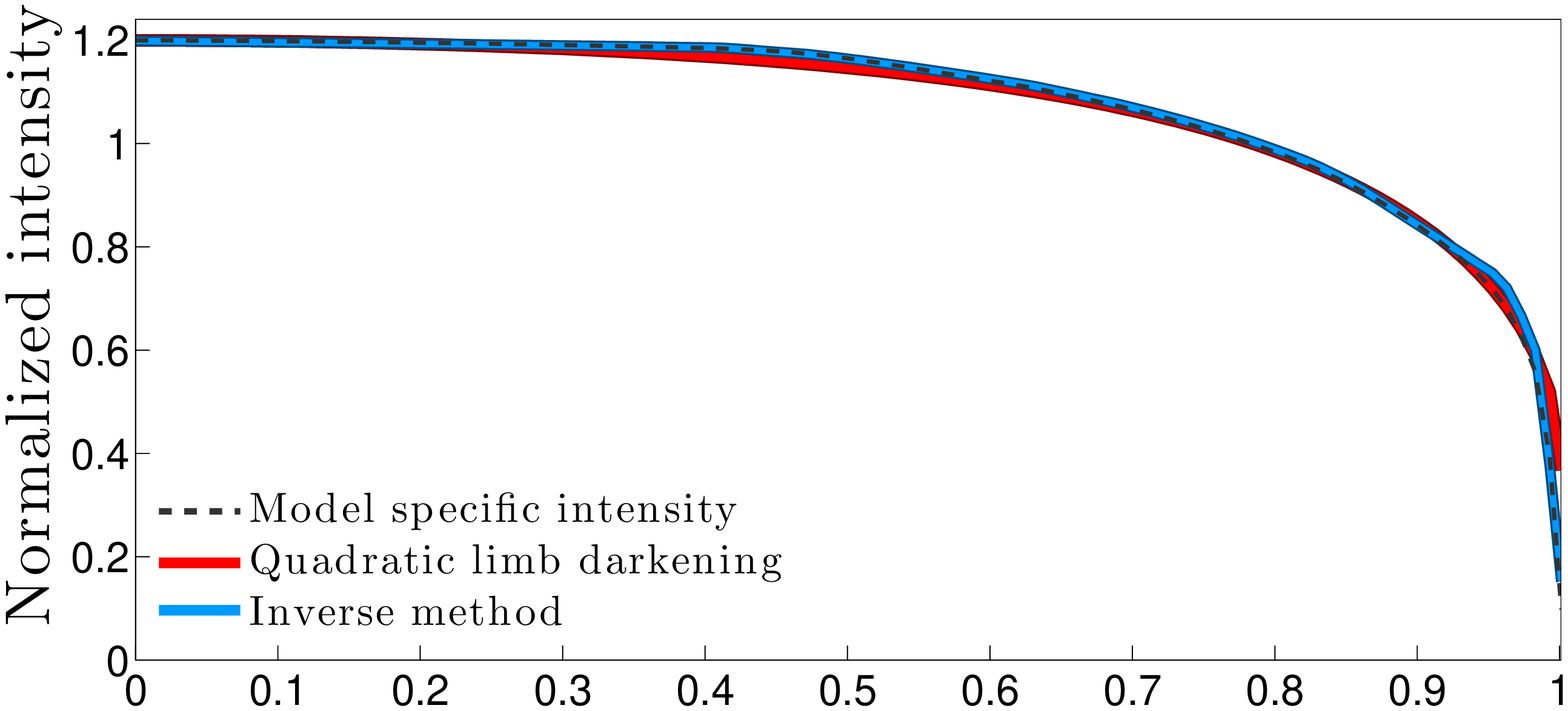}
	\includegraphics[width=\hsize]{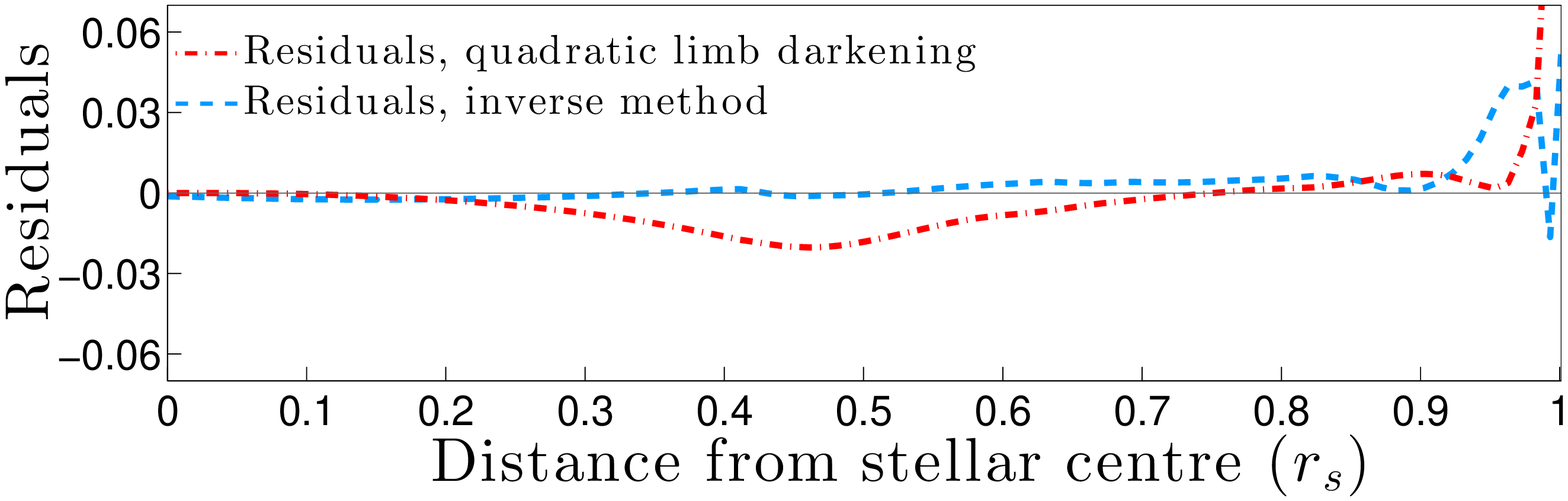}
	\caption{Comparison between inverse method and quadratic limb darkening, for recreating stellar center-to-limb brightness variations.\newline
		In black, specific intensity directly from the stellar model. Stellar model taken from case 2 in Table~\ref{planetPrameters} at $\lambda = 0.850 \mu m$.\newline
		In red, recreation of specific intensity using quadratic limb darkening, with coefficients $\gamma_1 = 0.21$ and $\gamma_2 = 0.45$.\newline
		In blue, recreation of specific intensity using the method presented in this paper.\newline
		Upper panel, direct comparison a continuum wavelength.\newline
		Lower panel, residuals, fitted specific intensity - model specific intensity.}
	\label{specIntVStest}
\end{figure}

\subsection{Finding the Regularization Parameter}  \label{detemineRegPara}
The chosen regularization parameter ($\alpha$) will have a large impact on the recovered solution. This is illustrated in upper and middle panel of Figure~\ref{optimalAlpha}, showing multiple solutions to the same data-set where the only difference is which $\alpha$ is used. This illustrates the importance of choosing a good value for $\alpha$. Too low $\alpha$ will result in a solution with large unphysical oscillations, and too high $\alpha$ removes real spectral features approaching wavelength-independent radius. In the lower panel of Figure~\ref{optimalAlpha}, estimates for reliability of derived results and goodness-of-fit to observations are shown as function of $\alpha$ (see Section~\ref{optimalregPara}). The minimum of the sum of these gives a value of $\alpha$ around 2600. Planetary radius recovered using this $\alpha$ is shown in the middle panel as a solid green line. Since this test was made using simulated data, the true value of the planetary radius was known. By calculating residuals between reconstructed and original planetary radius, the true optimal value of $\alpha$ could be computed ($\alpha_{optimal} \thickapprox 2900$ for this data-set). The planetary radius recovered using this optimal value is shown in the middle panel as a dashed turquoise line. This is the best possible solution for the given data. The solution using $\alpha$ obtained with the method from Section~\ref{optimalregPara} is almost identical to the solution with the true optimal value.

\begin{figure}
	\centering
	\includegraphics[width=\hsize]{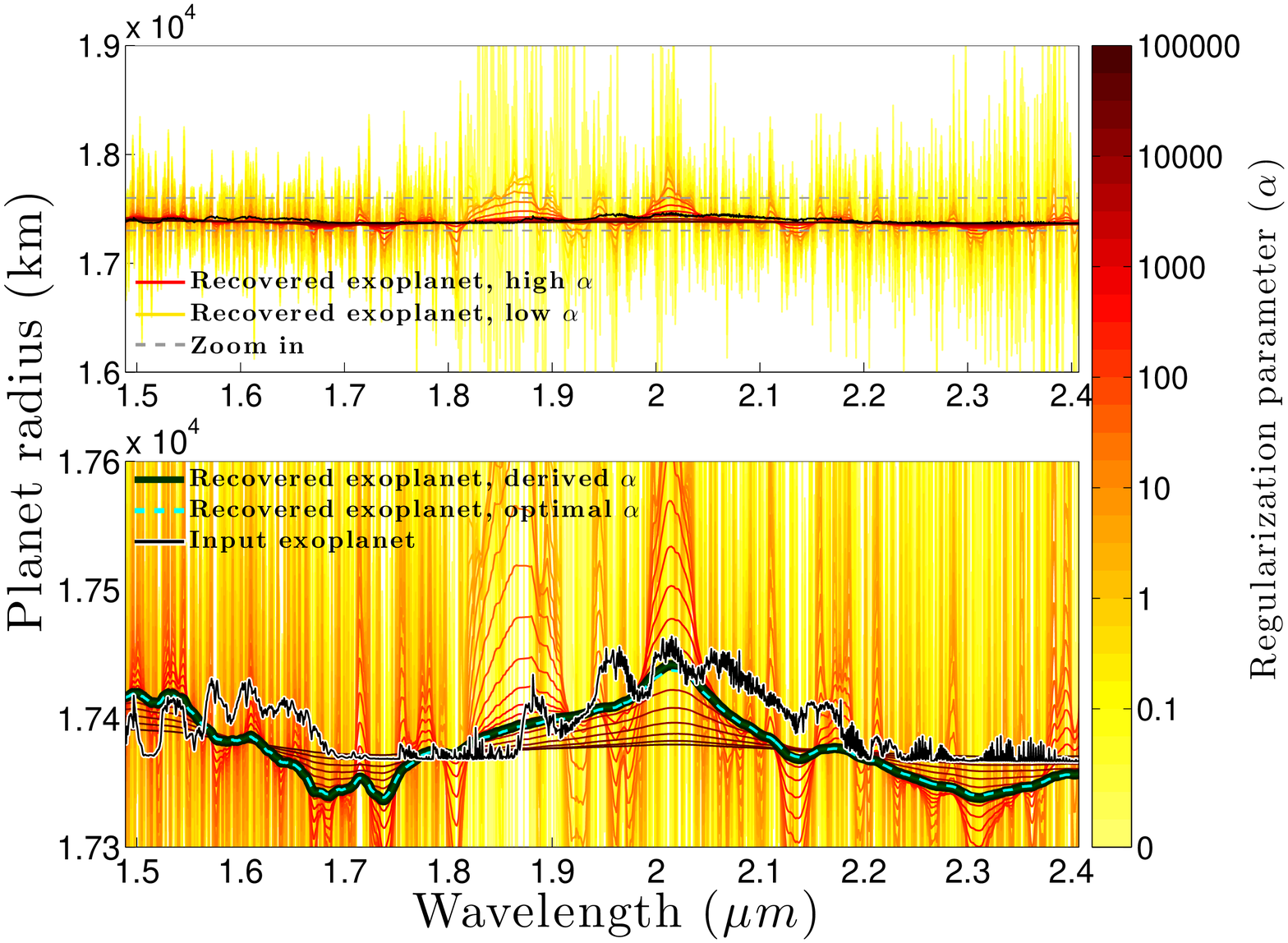}
	\includegraphics[width=\hsize]{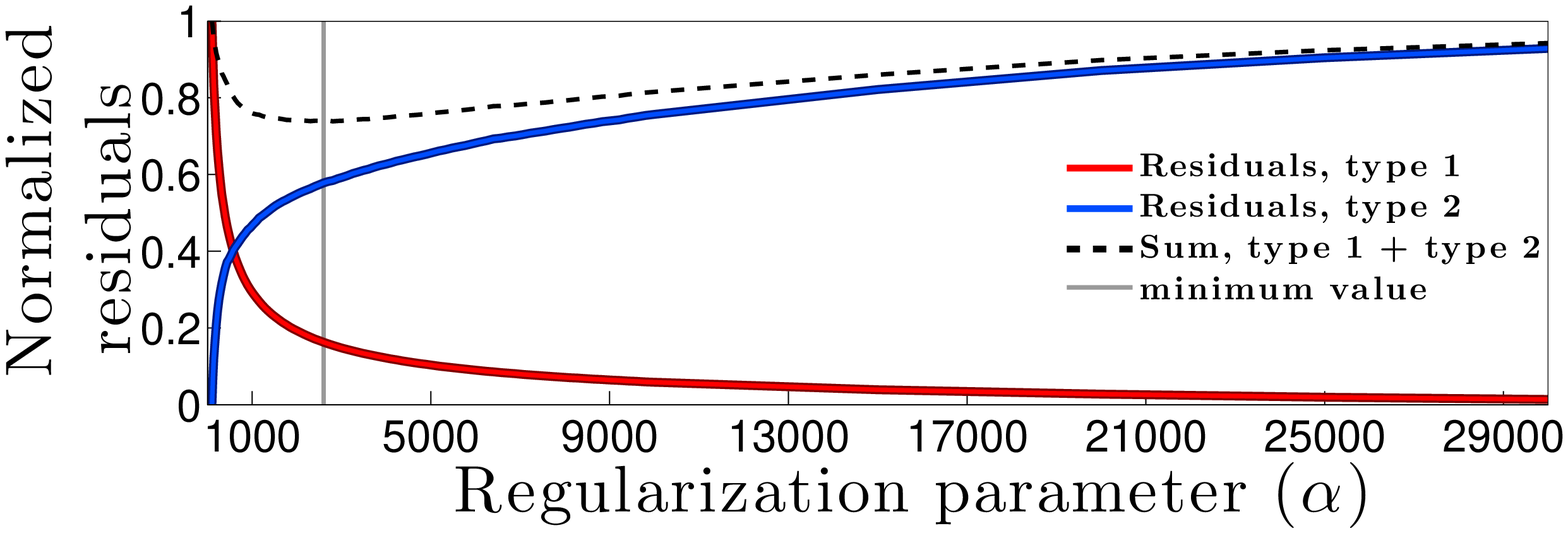}
	\caption{Impact of regularization parameter on recovered solution. \newline
		Upper panel: Recovered solutions are shown for regularization parameter ranging from $0$ to $10^{5}$. As regularization increases, the solution approaches its mean value. \newline
		Middle panel: Zoom in on the central parts of what is shown in the upper panel (region marked with dashed grey lines). The true planetary radius is shown in black. The solution using the derived regularization parameter from the method described in Section~\ref{optimalregPara} is shown as solid green line and the solution using the true optimal regularization parameter is shown as dashed turquoise line.\newline 
		Bottom panel: Normalized residuals as function of regularization parameter (see Section~\ref{optimalregPara} for definition of type 1 and type 2 residuals). The regularization parameter at the minimum value of the sum gives the value used to obtain the final solution (green line in the middle panel).}
	\label{optimalAlpha}
\end{figure}

\subsection{Specific Intensity Reconstruction} \label{SpecificIntensityReconstruction}
The proposed method can also be tested in a different manner. Instead of examining the accuracy of reconstructed planetary transmission, the accuracy of the derived stellar specific intensity can be assessed, since the specific intensity is fitted to the data simultaneously with planetary radius. Comparing the reconstructed specific intensity to stellar models can serve as test of the method, but it can also be of scientific interest as it can validate the correctness of stellar parameters and the quality of the model.

In Figure~\ref{specInteImag} we present the specific intensity at the center of the stellar disk, normalized to the average intensity. We show specific intensity derived from the stellar model used to synthesize observations and the one recovered from observations as described in Section~\ref{specificIntensity}, and filtered according to Equation~\ref{Tikho1}. The two versions of specific intensity are shown: In bright green, the recovered specific intensity from case 2 (Table~\ref{planetPrameters}), i.e. super-Earth transiting a small star viewed from surface of the Earth (in red). And in dark green, what can theoretically be achieved given good enough data. SNR in the latter case corresponds to more than 10 observed transits, and no telluric absorption was added to the synthetic observations. The reconstruction is in this case is recovered close to perfect, which shows that the proposed method can accurately reconstruct the specific intensity. Note, that the reconstruction is pure data processing not involving models of stellar or planetary atmospheres. In a more realistic case, the overall shape of the reconstructed specific intensity matches the model, but only the strongest and most prominent spectral features were recovered.

\begin{figure} [h]
	\centering
	\includegraphics[width=\hsize]{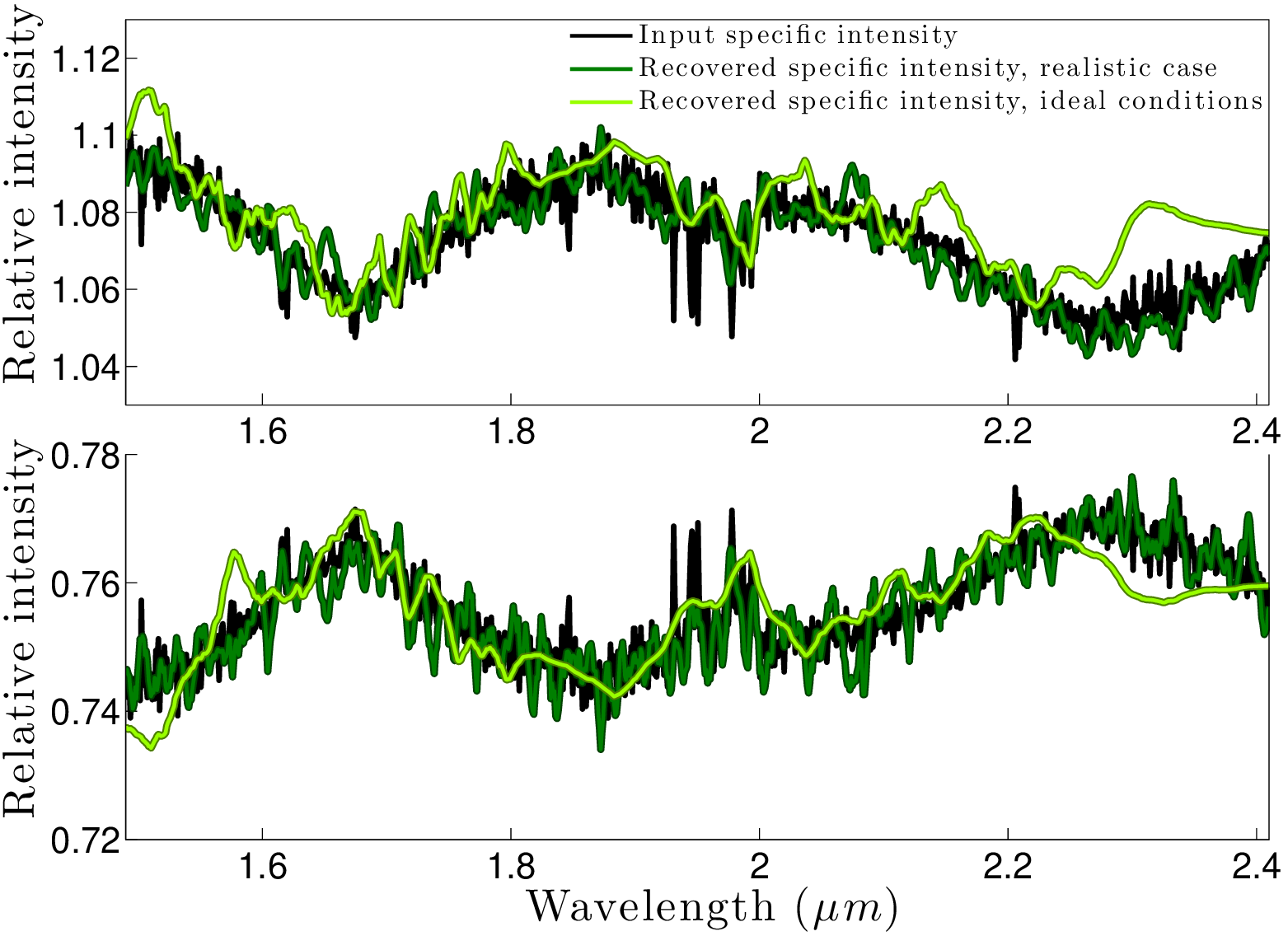}
	\caption{Recovered specific intensity compared to specific intensity taken directly from the model used to synthesize the observations. \newline
	Upper panel: Center of stellar disc. \newline
		Lower panel: Close ot stellar limb.}
	\label{specInteImag}
\end{figure}

\section{GJ 1214} \label{gj1214b} 
\subsection{Observations}
In this section we showcase the capabilities of the data analysis method by applying it to a set of transit observations of the super-Earth GJ 1214 b \citep{Charbonneau2009Natur}. GJ 1214 b is a super-Earth in close in orbit around a M-type dwarf star. The small size of the star (0.21 $R_{\odot}$) provides a high planetary to stellar radius ratio. This combined with the brightness of the star makes GJ 1214 b an ideal target for transit spectroscopy which has been extensively observed in the past. Note that we do not aim to present new scientific results on the GJ 1214 b system, only demonstrate that the method is able to deliver high-quality results using actual data. We retrieved observations of 9 transits from the ESO archive. All observations were made with the same instrument, FORS2 at the VLT \citep{Appenzeller1998Msngr}, using the low resolution multi-object spectroscopy mode. During each transit, GJ 1214 and 5 reference stars were observed simultaneously. See \citet{Bean2010Natur, Bean2011ApJ, Berta2011ApJ} and Table~\ref{GJ1214dates} for more details

\begin{table}[h]
	\caption{Transit observations of GJ 1214b}
	\centering
	\begin{tabular}{lll}
		\hline \hline
		Date & ESO program ID & wavelength region\\
		\hline
		2010-04-29 & 284.C-5042(B) & 0.71 - 1.02 $\mu m$ \\ 
		2010-05-18 & 285.C-5019(A) &  0.71 - 1.02 $\mu m$ \\ 
		2010-07-22 & 285.C-5019(C) &  0.71 - 1.02 $\mu m$ \\ 
		2011-07-03 & 087.C-0505(A) &  0.57 - 0.88 $\mu m$ \\ 
		2012-06-13 & 089.C-0020(G) &  0.73 - 1.03 $\mu m$ \\  
		2012-07-02 & 089.C-0020(H) &  0.73 - 1.03 $\mu m$ \\ 
		2012-07-21 & 089.C-0020(I) &   0.73 - 1.03 $\mu m$ \\ 
		2012-07-29 & 089.C-0020(D) & 0.73 - 1.03 $\mu m$ \\  
		2012-08-17 & 089.C-0020(B) &  0.73 - 1.03 $\mu m$ \\  
		\hline
	\end{tabular}
	\label{GJ1214dates}
\end{table}

\subsection{Data Reduction} \label{GJ1214_DataReduction}
Initial data reduction was carried out in a similar fashion as described by \citet{Bean2010Natur}. Visual representation of the data reduction of a single transit observation is shown in Fig \ref{GJReduction}. The upper panel shows GJ 1214 spectra after initial reduction (flat fielding, wavelength calibration etc). At this stage only stellar and telluric lines are visible. GJ 1214 spectra were then divided by the SNR-weighted average of all reference stars. This was done for each spectral pixel and exposure independently, using the corresponding wavelength and exposure of the reference stars. This step removes temporal changes in brightness that is affecting all stars (predominately changes in the telluric absorption). This step is shown in the middle panel. After this, each wavelength channel was normalized by the average out-of-transit flux in the same wavelength, shown in the lower panel. Temporal changes of unknown origin could however still be detected in the out-of-transit flux of GJ 1214. These were removed by fitting a low-order polynomial to the upper envelope of the wavelength-averaged light curve and normalizing the flux to out-of-transit levels, see Figure~\ref{RemResi}.

We looked for signatures of stellar spots in white light flux. Two probable spot-crossing events were detected, and the corresponding exposures were removed before further analysis. Data with exceptionally low SNR (bad pixels, strong telluric absorption etc) were also removed from the data-set. Not counting removed data from non-overlapping spectral coverage, a total of around 5\% of pixels where removed during this process.

Once telluric transmission had been removed from all transit observations, there were still inconsistencies in transit depth between observations taken at different nights. This can be explained by a known stellar variability of 1\% \citep{Berta2011ApJ}. To account for this effect, the wavelength averaged transit depth (in the overlapping spectral region) was scaled to the same average transit depth for all observations. This depth was set to the median transit depth of all observed transits, which required rescaling of transit depth by 0.7\% in the most extreme case (which also happen to be the transit with the lowest SNR) and less than 0.2\% in all other cases, consistent with the stellar variability of 1\%. 

\begin{figure}
	\centering
	\includegraphics[width=\hsize]{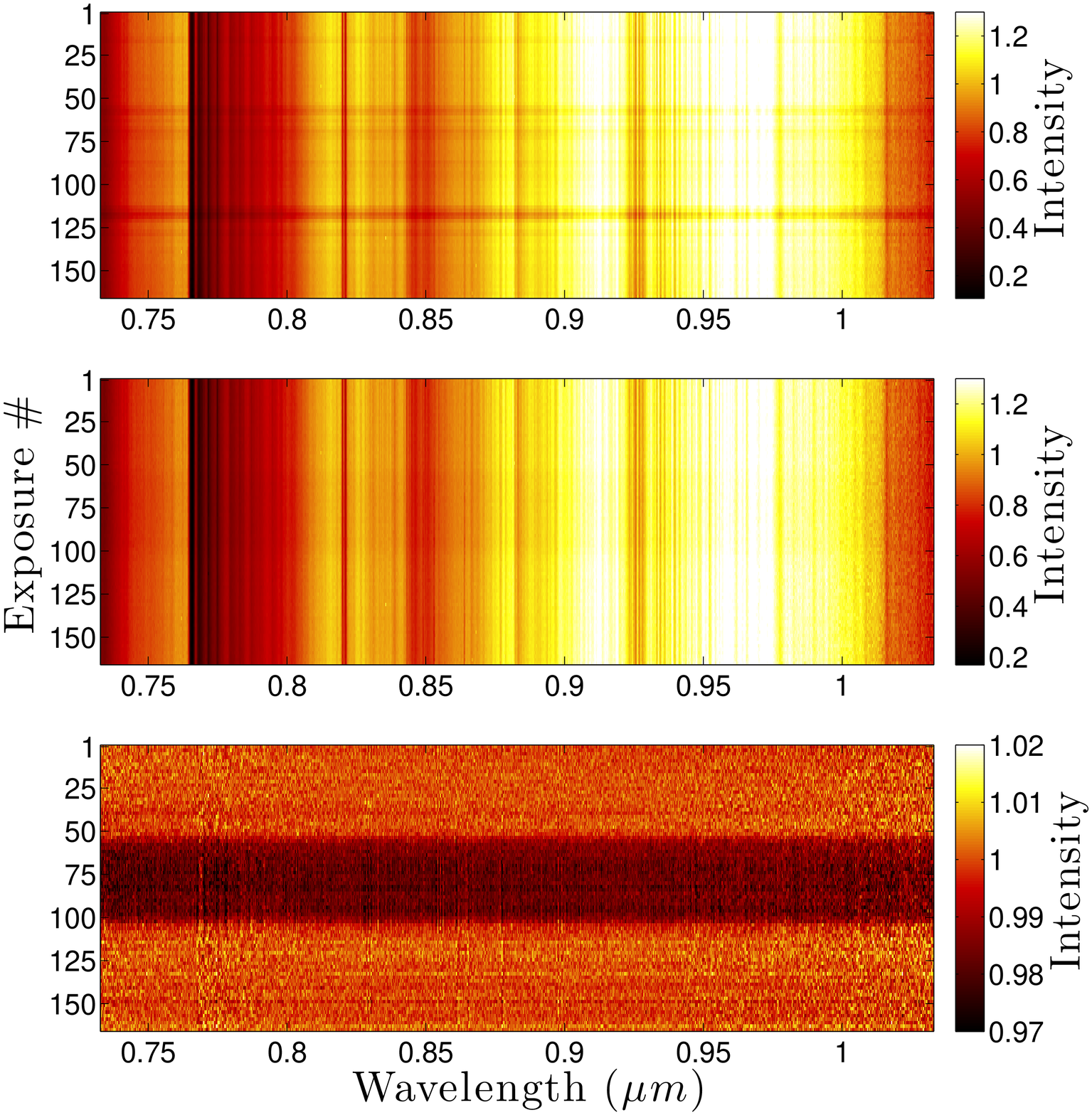}
	\caption{Reduction procedure for FORS2 observation of GJ 1214 during a transit event. The vertical axis shows the change in stellar flux over time.\newline 
		Upper panel: Observations after initial reduction (flat fielding, bias, wavelength calibration etc). At this scaling all visible lines have telluric or stellar origin. Horizontal lines shows sudden decrease in telluric transmission. \newline
		Middle panel: Observations after division by reference stars, removing temporal changes in telluric transmission. \newline
		Lower panel: Each wavelength channel has been normalization by the flux in out-of-transit exposures. Results shows typical transit light curves with average transit depth in agreement with previously determined size of the planet.}
	\label{GJReduction}
\end{figure}

\begin{figure}
	\centering
	\includegraphics[width=\hsize]{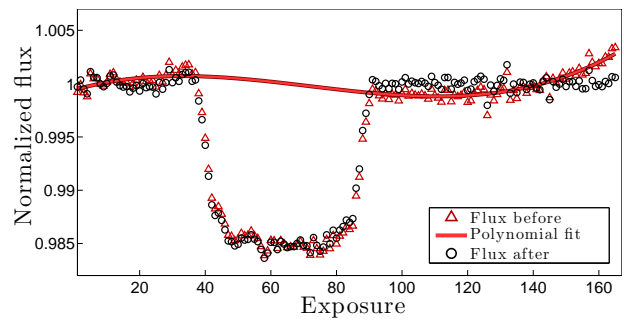}
	\caption{Example of removal of telluric residuals from transit observations of GJ 1214b. A third degree polynomial is fitted to wavelength averaged flux of GJ 1214. By normalizing with this polynomial the wavelength independent residuals are removed. Note that the two bumps that can be seen here are not the two removed spot-crossing events.}
	\label{RemResi}
\end{figure}

\subsection{Data Analysis} \label{GJ1214_DataAnalysis}
In order to take full advantage of the data analysis method, we combine all 9 transit observations into a single wavelength dependent transit light curve. First, all spectra were interpolated onto a common wavelength grid, removing all non-overlapping wavelength regions (with the exception of the observation from 2011, since this had very different spectral coverage, see Table~\ref{GJ1214dates}). Exposures were then sorted by the planet's orbital phase (i.e. the position of the planet during the exposure). This creates a single transit light curve for each wavelength channel, with up to 550 exposures taken during transit (see Figure ~\ref{combGJ}). Removed bad data and missing data (due to differences in observed wavelength region) is shown in turquoise. The data analysis method described in Section~\ref{analysisMethod} was then applied to these combined light curves. 

\begin{figure}
	\centering
	\includegraphics[width=\hsize]{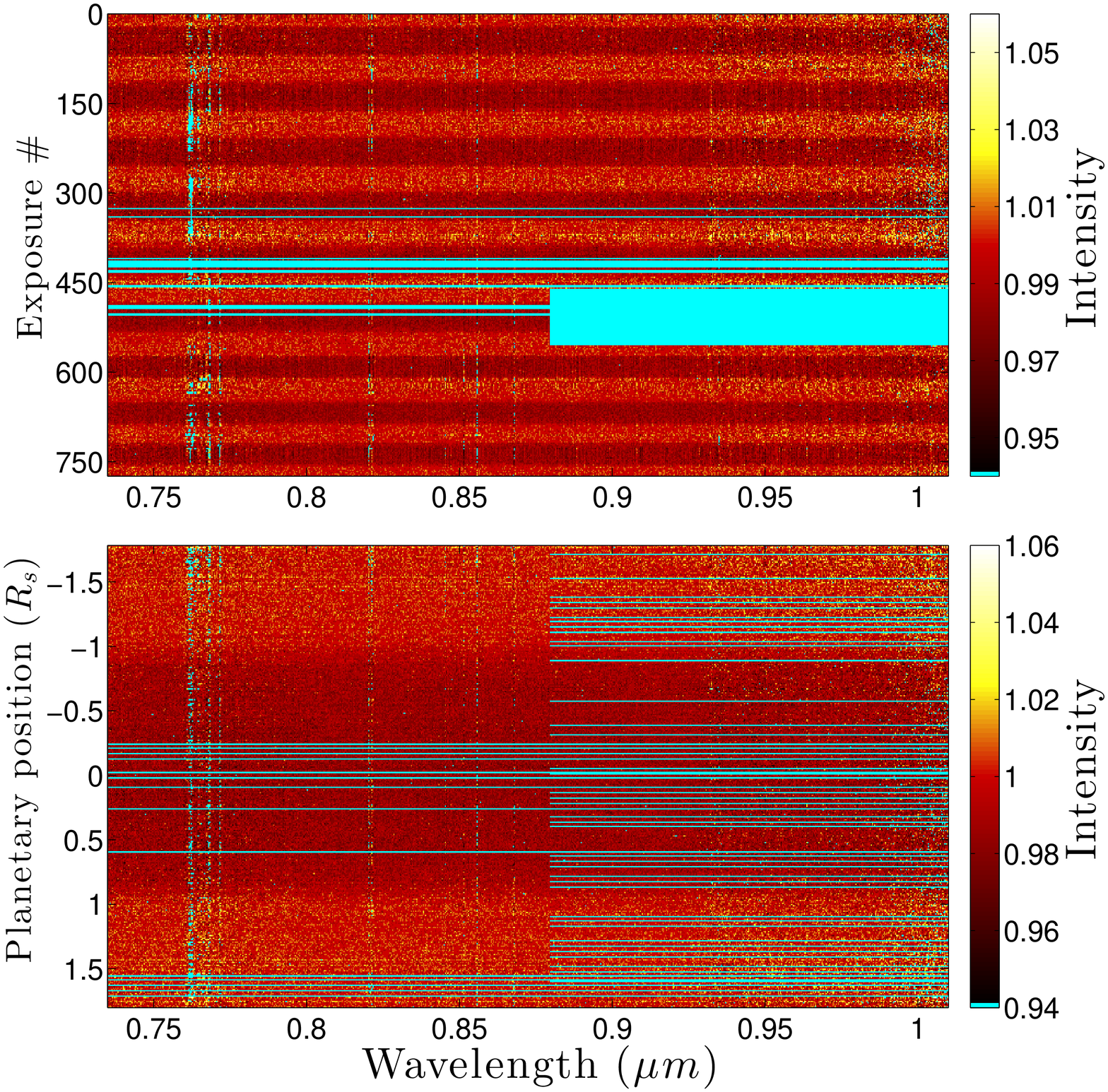}
	\caption{Visual representation of procedure for combing observation of 9 transits of GJ 1214 b. Data points marked with turquoise are either very noisy or missing in the data-set, and were ignored during data analysis.\newline
		Upper panel: All transits were interpolated onto a common wavelength grid. \newline
		Lower panel: Data sorted by the exoplanet's orbital phase, creating a combined transit brightness curve.}
	\label{combGJ}
\end{figure}
We also used a different approach to recovering exoplanet transmission spectrum. We analysed each transit separately, and then combined the recovered exoplanet transmission spectra (rather than first combining all observations into a single "transit"). Due to smaller observational constrains this approach forces us to use higher regularisation parameter effectively reducing spectral resolution. For larger planets individual transit should have enough SNR to reconstruct transmission spectrum providing a consistency check and allowing test for atmospheric variability.  

\subsection{Results - Exoplanet Transmittance}
The transmission spectrum of GJ 1214 b that we recovered is shown in Figure~\ref{GJ1214b_spec} as a solid red line. The 1$\sigma$ confidence intervals (shown in dashed red) were estimated using bootstrap analysis, reshuffling the order of wavelengths points before applying the regularizing filter to get the typical distribution of data points for the given data-set and regularization. Note that this is the confidence intervals for the regularized spectrum, the true non-regularized transmission spectrum is not expected to be restricted by these limits. The recovered spectrum from each individual transit observation is also shown in the same figure, as dotted lines, and the average of these are shown as a solid blue line. The latter is very similar to the recovered solution from the combined transit (red line), supporting our confidence in the results. Due to the small differences in transit depth between transit events, all spectra are normalized by their average value (which is $2.699 R_{\oplus}$ for the combed case, i.e. the red line). Finally the average telluric transmittance and stellar flux are shown in order to demonstrate that the recovered solution bears no resemblance to either of them and thus it is unlikely to be affected by deficiencies in data rectification. 

The recovered spectra using all transits are mostly featureless, in agreement with what has been measured in many previously studies \citep{Bean2010Natur, Bean2011ApJ, Carter2011ApJ, Croll2011, Crossfield2011ApJ, desert2011ApJ, Berta2012ApJ, Murgas2012A&A, deMooij2012AA, deMooij2013ApJ, Colon2013ApJ, Fraine2013ApJ, Narita2013ApJ, Teske2013MNRAS, Wilson2014MNRAS, Caceres2014A&A, Kreidberg2014Natur}. A decrease of planetary radius with longer wavelengths is however detected, which could be explained by a thick cloud cover and Rayleigh scattering in a hydrogen dominated metal-poor atmosphere above these clouds. As this paper is focusing on the methodology of data analysis for transit spectroscopy, we do not dive into a more detailed interpretation of these results.

\subsection{Results - Stellar Specific Intensity}
The specific intensities reconstructed simultaneously with the planetary transmittance are subjected to filtering (see Section~\ref{SpecificIntensityReconstruction}) before comparing to specific intensities derived from model atmosphere. For comparison of flux spectra we median-combine flux spectra of GJ 1214, which creates a single spectrum with high SNR. We did the same with the brightest reference star, from which the telluric transmittance was obtained. Telluric was then removed from the GJ 1214 flux spectrum, resulting in the stellar flux of GJ 1214. The numerical model with good fit to this was then chosen from a grid of models. Model parameters are shown in Table~\ref{gj1214stellarPara}, and do not differ significantly from the values derived by \citet{Harps2013A&A}. Both MARCS models \citep{Gustafsson2008} and PHOENIX models \citep{Husser2013A&A} were tried (see upper panel of Fig \ref{GJ1214specIntRec}). Spectral resolution was adjusted to match the resolution of the observed data. We note a very good consistency in reproduction of spectral features and the center-to-limb variation. The main discrepancy between models and observations in the flux spectra is the absorption feature around 1.0 $\mu m$ (likely $FeH$ and $CrH$), where MARCS models underestimate the absorption and PHOENIX models overestimate it (the two model codes are using different partition function and dissociation energy data for these molecules). Finding other discrepancies between observations and models is difficult at this resolution. The obtained specific intensity is however less sensitive to instrumental effects and imperfect telluric removal, because it is a differential measurement: Using the intensity at a given point on the stellar disk relative to the average intensity of the entire disk removes most of instrumental effects and residual telluric features emphasizing the contrast between the center of disk and the limb as a function of wavelength. Comparison of the reconstructed and modeled specific intensity is shown in middle and lower panel of Fig \ref{GJ1214specIntRec}.

Overall the reconstructed specific intensity matches the models, i.e. the predicted contrast difference between the center and the limb of the disk resembles measurement. The fit in regions with strong telluric absorption is noticeably worse, as can be seen clearly around the $O_2$ band at 0.76 $\mu m$ and the $H_2O$ band around 0.95 $\mu m$. Strong spectral features also appear weaker. This  is expected due to the smoothing effect of regularization. With decreasing regularization (shown as gray line in Fig \ref{GJ1214specIntRec}), the depth of the strong lines is reconstructed better but the reconstruction quickly becomes noisy overall. 

Specific intensity can be used to validate stellar models, since relative specific intensity hold information about opacities as function of depth. The contrast ratio between the center and the limb of the disk is an indirect measurement of how deep into the star the detected radiation is originating from. The derived specific intensity for this star indicates that we are unable to model $CrH$ opacities as function of height correctly. Both the $CrH$ band around 0.88 $\mu m$ and 1.02 $\mu m$ show significant difference between the observed intensity and what is predicted by models. 

\begin{figure}
	\centering
	\includegraphics[width=\hsize]{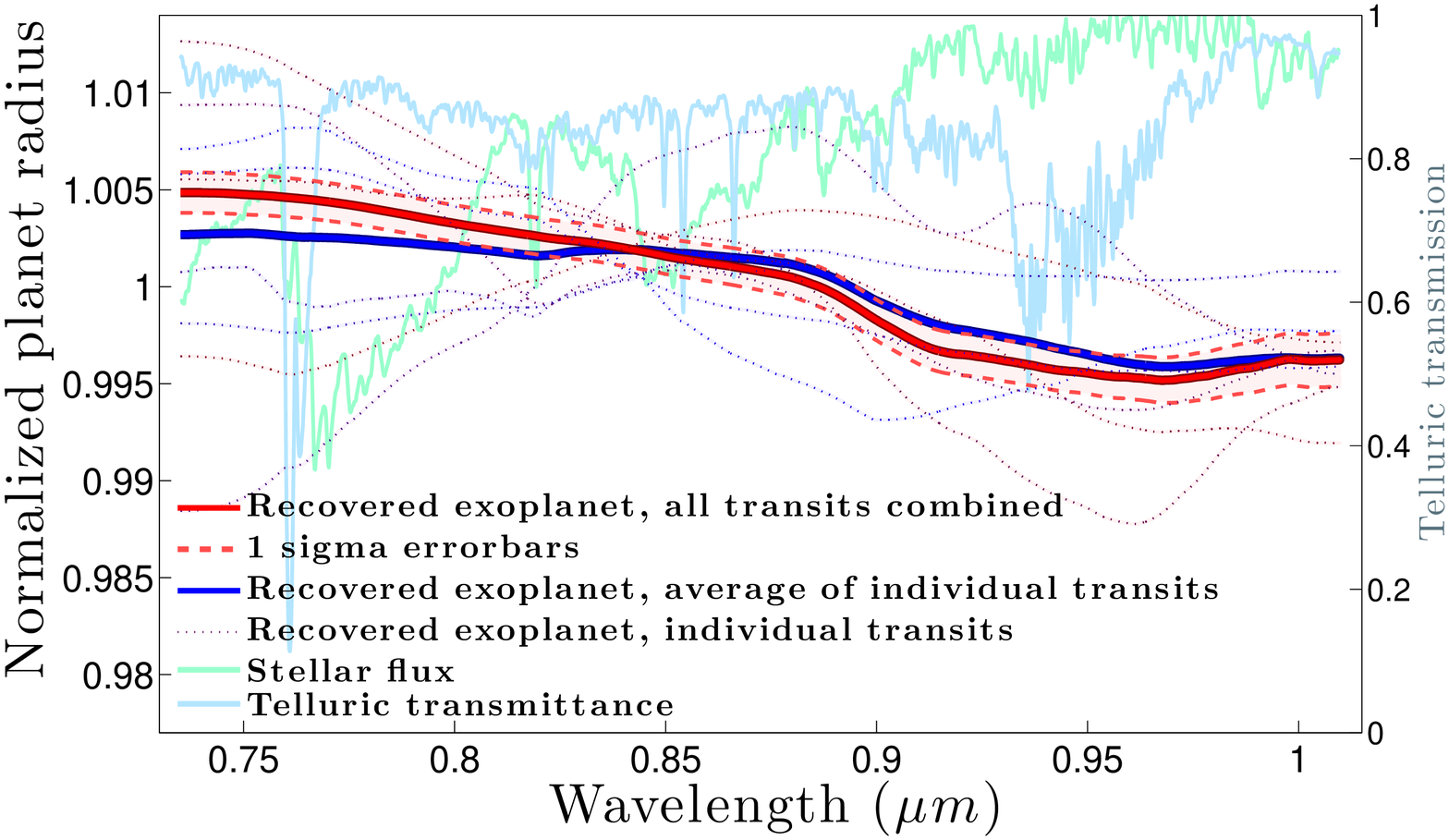}
	\caption{Recovered transmission spectrum of GJ 1214 b. In red, the solution to the combined transit light curve, with $1\sigma$ confidence intervals (for the regularized spectrum) in dashed red. As dotted lines, the recovered transmission spectrum from each individual transit, and the average of these as solid blue. \newline
		No significant spectral features are detected, except an overall decrease of planetary radius with longer wavelengths.}
	\label{GJ1214b_spec}
\end{figure}

\begin{table}[h]
	\caption{Stellar parameters for GJ 1214. Literature values from \citet{Harps2013A&A}}
	\centering
	\begin{tabular}{lccc}
		\hline  \hline
		& MARCS & PHOENIX & Literature \\
		\hline
		$\mathrm{T_{eff}}$ (K) &2990 &3000 & 3026\\
		$\mathrm{\frac{Fe}{H}}$ (dex) & +0.50 & +0.50 & +0.39\\
		$\mathrm{log g_s}$ (cgs) & 4.95 & 5.0 & 4.994\\
		\hline
	\end{tabular}
	\label{gj1214stellarPara}
\end{table}

\begin{figure}
	\centering
	\includegraphics[width=\hsize]{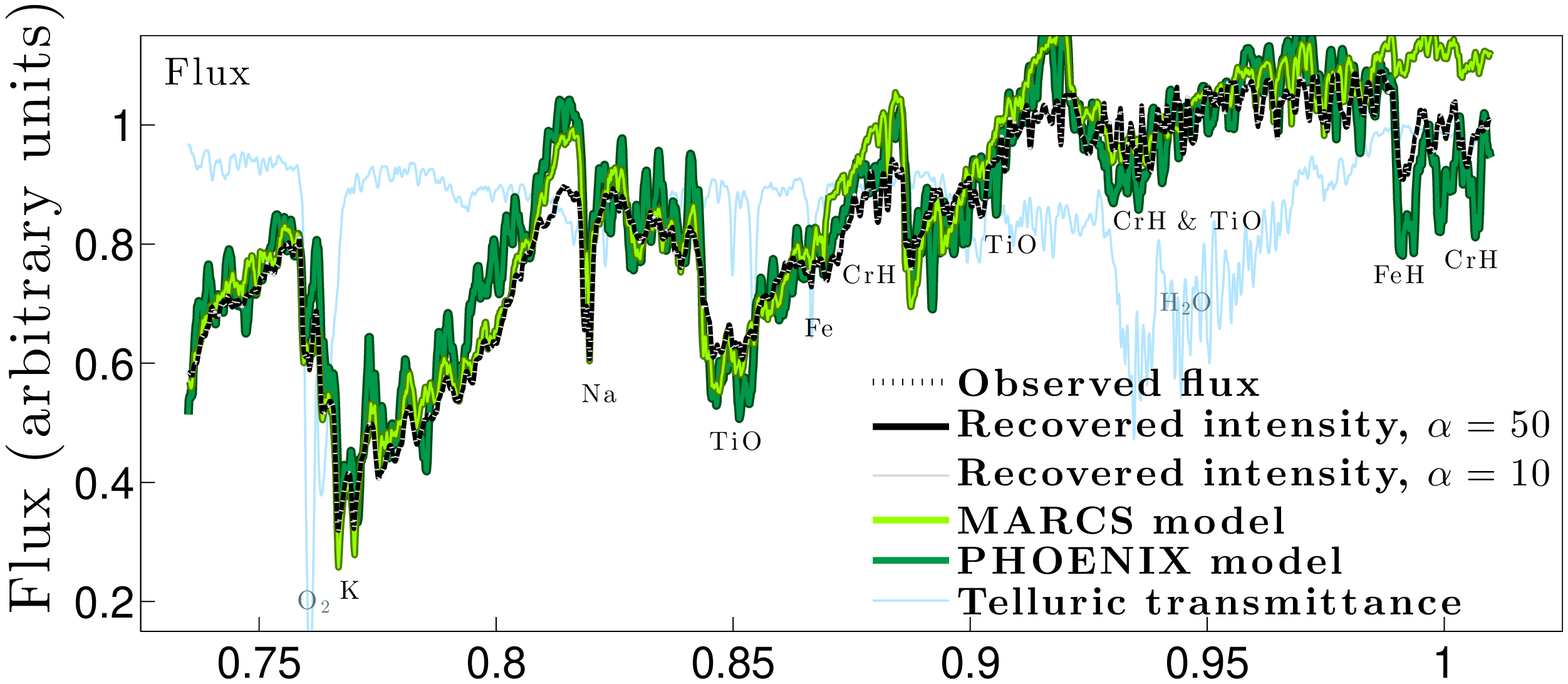}
	\includegraphics[width=\hsize]{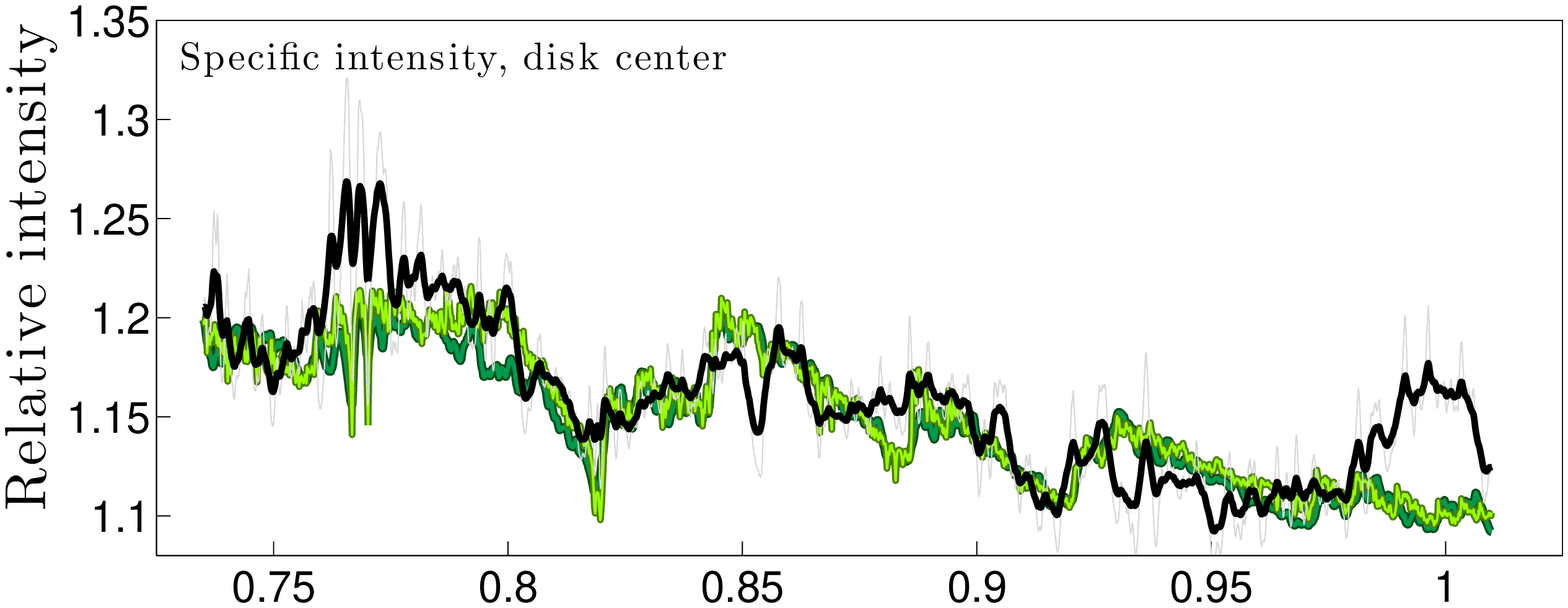}
	\includegraphics[width=\hsize]{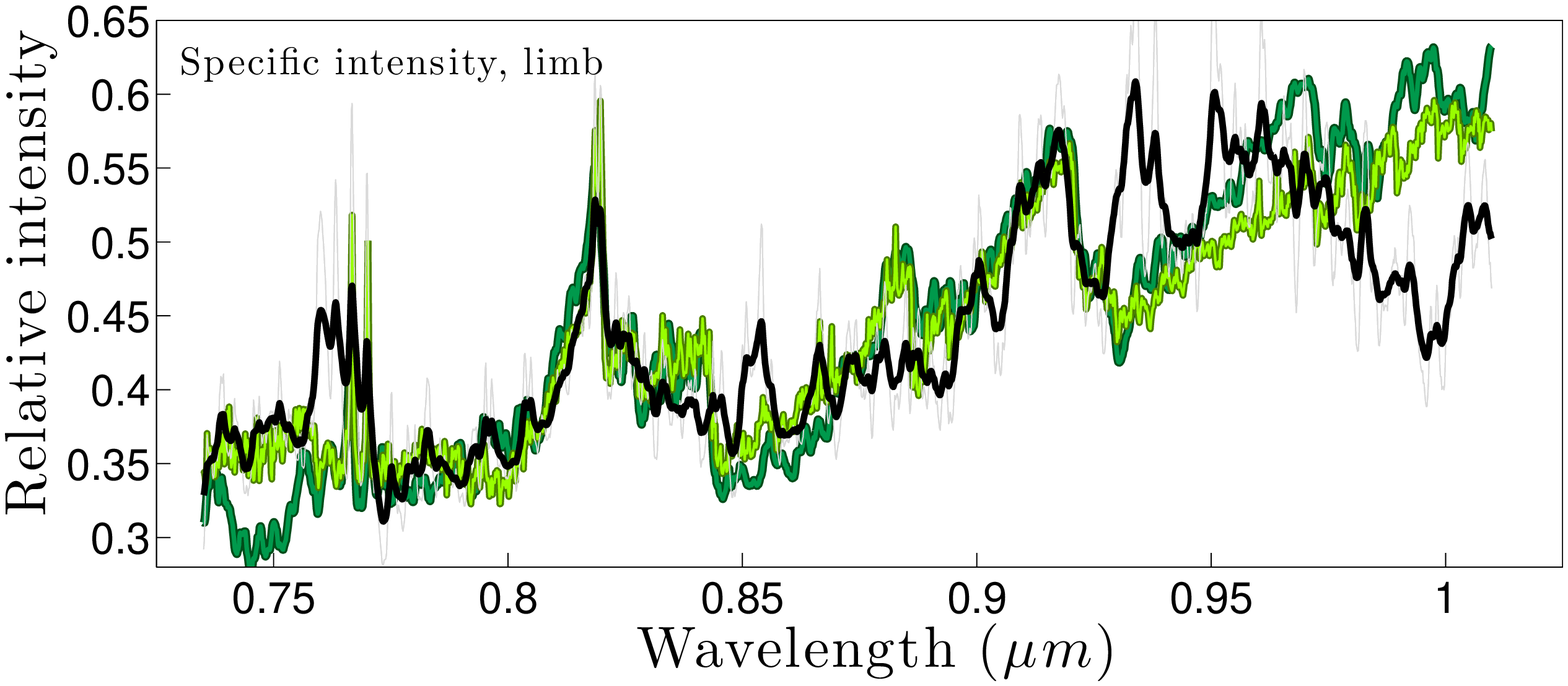}
	\caption{Reconstructed specific intensity of GJ 1214, plotted alongside stellar models.\newline
		Upper panel: Flux spectrum from GJ 1214 and the best fitting models.\newline
		Middle panel: Specific intensity close to the center of the disk, $\mu = 0.9$. \newline
		Lower panel: Specific intensity close to the limb of the disk, $\mu = 0.1$. \newline
		Note that the specific intensity is given relative to the average specific intensity across the entire disk. Low relative intensity in the center of the disk and high relative intensity at the limb of the disk indicates a flat intensity distribution, and a higher value at the center and lower at the limb indicates a steeper contrast difference.}
	\label{GJ1214specIntRec}
\end{figure}

\section{WASP-49} \label{wasp49}
\subsection{Observations}
WASP-49 b is a hot-Jupiter in close orbit (67 hour period) around a G6 type dwarf star \citep{Lendl2012AA}. We applied the data analysis method on observations of this system during a three transit events (see Table~\ref{wasp49dates}). Observations were carried out with the same instrument (FORS2 at VLT), where the target (WASP-49) and three reference stars were observed simultaneously using the multi-object spectroscopy MXU mode. Detailed description on observations can be found in \citet{Lendl2016AA}.

\begin{table}[h]
	\caption{Transit observations of WASP-49 b}
	\centering
	\begin{tabular}{lll}
		\hline \hline
		Date & ESO program ID & wavelength region\\
		\hline
2012-12-05 & 090.C-0758(B)  &  0.73 - 1.03 $\mu m$ \\ 
2013-01-14 & 090.C-0758(C)  &  0.73 - 1.03 $\mu m$ \\ 
2013-02-07 & 090.C-0758(D) &  0.73 - 1.03 $\mu m$ \\ 
		\hline
	\end{tabular}
	\label{wasp49dates}
\end{table}

Data reduction and data analysis was carried out in the same way as for GJ 1214 (see Section~\ref{GJ1214_DataReduction} and \ref{GJ1214_DataAnalysis}). Combination of all transit observations into a single wavelength dependent transit light curve is shown in Figure~\ref{combWASP}. Many exposures suffered from over-saturated pixels, which lead to a large fraction of pixels being removed before analysis (around 12\%). This is marked in Figure~\ref{combWASP} with turquoise.

\begin{figure}
	\centering
	\includegraphics[width=\hsize]{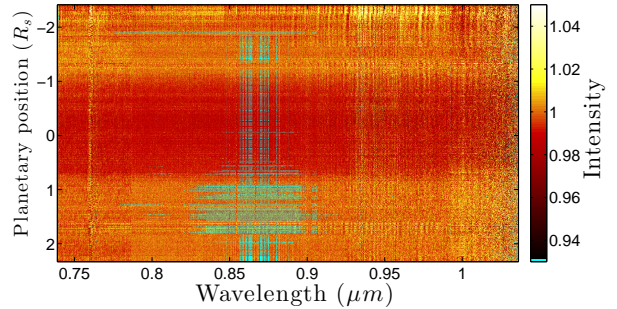}
	\caption{The observational data from three transits of WASP-49 b are here ordered by wavelength and orbital phase, creating a single wavelength dependent light curve. Data points marked with turquoise are removed from the data-set (bad pixels, over-saturated pixels etc) before analysis.}
	\label{combWASP}
\end{figure}

\subsection{Results - Exoplanet Transmittance and Stellar Specific Intensity}
The quality of observations is insufficient to detect any significant spectral features in the recovered exoplanet transmission spectrum as seen in Figure~\ref{atmoWASP}. The spectrum is mostly flat with an overall trend of higher opacity at longer wavelengths and a sharp decrease in opacity around 0.75 $\mu m$. Again, this paper focuses on the data analysis method rather than the interpretation of derived results so we will not attempt to establish the physical properties of the atmosphere of WASP-49 b. 

Recovered stellar specific intensity of WASP-49 is shown in Figure~\ref{intenityWASP}, along with a PHOENIX model and a MARCS model with similar stellar parameters (see Table~\ref{WASP49stellarPara}). The overall trend of the recovered specific intensity is in good agreement with the model, and most spectral lines are recovered. We find that the agreement with the model is noticeably worse beyond 0.95 $\mu m$. We believe this is a result of poor data quality in parts of this wavelength region due to strong telluric absorption.

\begin{figure}
	\centering
	\includegraphics[width=\hsize]{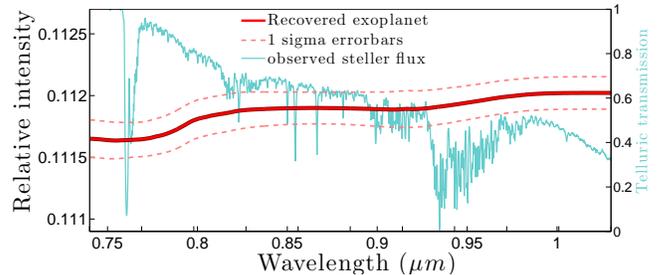}
	\caption{Exoplanetary transmission spectrum of WASP-49 b shown in red. No significant spectral features are detected. Stellar flux, as seen through the telluric atmosphere shown in blue.}
	\label{atmoWASP}
\end{figure}

\begin{figure}
	\centering
	\includegraphics[width=\hsize]{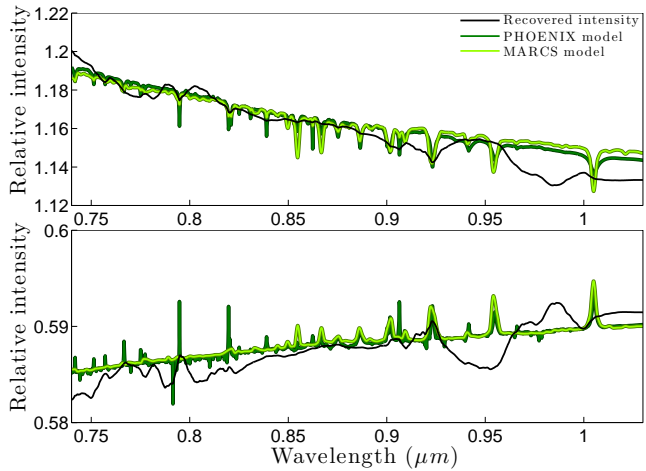}
		\caption{Recovered specific intensity of WASP-49 shown in black, plotted over MARCS and PHOENIX models with similar stellar parameters, in bright and dark green respectively . The general trend, as well as most strong special features matches the models.\newline
	Upper panel: Center of stellar disc. \newline
		Lower panel: Close ot stellar limb.}
	\label{intenityWASP}
\end{figure}

\begin{table}[h]
	\caption{Stellar parameters for WASP-49. Literature values from \citet{Lendl2016AA}}
	\centering
	\begin{tabular}{lccc}
		\hline  \hline
		& MARCS & PHOENIX & Literature \\
		\hline
		$\mathrm{T_{eff}}$ (K) & 5600 & 5600 & 5602\\
		$\mathrm{\frac{Fe}{H}}$ (dex) & -0.25 & -0.25 & -0.23\\
		$\mathrm{log g_s}$ (cgs) & 4.5 & 4.5 & 4.406\\
		\hline
	\end{tabular}
	\label{WASP49stellarPara}
\end{table}

\section{CONCLUSIONS} \label{Conclusions}
The data analysis method presented in this paper allows for robust recovery of exoplanetary transmission spectra, without any model dependences. The telluric transmittance is taken directly from observations of telluric standard stars and the stellar specific intensity is self-consistently reconstructed simultaneously with the "effective" exoplanetary radius. We have found that our method for fitting specific intensity is not more dependent on accurate orbital parameters or treatment of stellar spots that standard methods based on quadratic limb darkening equations. A regularization-based filtering technique helps finding a compromise between spectral resolution and SNR, for example, in regions dominated by telluric absorption. This enables us to extract more information form a given data-set than standard methods, by keeping a high spectral resolution in regions where data quality allows for this and reducing resolution in regions where this is necessary in order to obtain reliable results. 

Planetary atmospheric models is still required for the interpretation of the results in terms of physical conditions (temperature, pressure and detailed chemical composition) but detection of chemical species can be done already from the reconstructed transmission spectrum from the shape and position of absorption features. By using a grid of planetary models with a range of chemical abundances and temperature/pressure structures, the model with the best fit to the recovered transmission spectrum could be found. The same regularizing filter should be applied to model transmission to match variable spectral resolution of the processed observations. 

The data analysis method described in this paper requires a special set of observations: a sequence of spectra of a star during a transit event using short integration times (1-2 minutes) complemented by simultaneous spectra of reference stars for tracing temporal changes in telluric transmittance (for ground-based observations) and instrumental setup. Optimal instruments for this type of observations are low resolution multi-object spectrographs. The method could also be applied to space based observations, and if so, the need for monitoring reference stars diminishes and thereby a single-object spectrograph may be sufficient. Space-based observations also offer the best chances to detect water in exoatmospheres, which due to strong absorbing water in the telluric atmosphere is difficult to archive from the ground. 

When selecting targets for observations of this type, the most critical parameter is the relative size of the exoplanet to its host star. Small stars and large exoplanets with extended atmospheres are ideal. Characterization of gas giants transiting small stars requires considerably less observing time than characterization of rocky planet transiting larger stars. 

The observations discussed in this paper could be efficiently collected with the JWST NIRSpec instrument (jwst-docs.stsci.edu/display/JTI/). Space observations alleviate the problems associated with telluric contamination and make observations of any transit possible. NIRSpec bright object time-series (BOTS) mode uses the 1.6” × 1.6” fixed slit aperture that is optimized for exoplanet transit observations requiring stable observing condition and high photometric precision time-series spectroscopy. “High" resolution (R=2700) configuration offer nearly continuous wavelength coverage of four spectral regions: 0.7-1.27, 0.97-1.89, 1.66-3.17 and 2.87-5.27 $\mathrm{\mu}$m, containing a range of interesting molecular bands including O$_2$, H$_2$O, CH$_4$, CO, CO$_2$ and others.

The limiting K magnitudes for these modes are around 7 for M-dwarfs and 8 for solar-type stars - comfortably above the targets considered in this paper (Sections \ref{gj1214b} and \ref{wasp49}). High sensitivity of the BOTS mode will allow taking over 210 exposures for GJ 1214 and over 220 exposures for WASP-49 during each transit reaching a SNR close to 200. The numbers obtained using the JWST exposure time calculator demonstrate the amazing efficiency of collecting low-resolution transit spectroscopy data with NIRSpec. The exquisite quality of NIRSpec spectra will be sensitive not only to the properties and size of exoplanetary atmosphere but also to the deficiencies of stellar and planetary models involved in the analysis. Therefore, we believe that separating the derivation of specific intensities and the transmittance spectrum of a planet from model-dependent interpretation is a more robust strategy for analyzing JWST exoplanet spectra.

\begin{acknowledgements}
	The authors would like to thank Bengt Edvardsson and Kjell Eriksson for a major help with exploring molecular opacities in MARCS models.\\
	We also acknowledge the ESO archive for providing a complete data-set for VLT FORS2 observations of GJ 1214 b (284.C-5042(B), 285.C-5019(A), 285.C-5019(C), 087.C-0505(A), 089.C-0020(G), 089.C-0020(H), 089.C-0020(I), 089.C-0020(D), 089.C-0020(B)) \\
	and WASP-49b (090.C-0758(B), 090.C-0758(C), 090.C-0758(D)).
	\\
\end{acknowledgements}

\section{APPENDIX} \label{appendix}
In this section we describe some of the methods used in numerical implementation of the data analysis presented in Section \ref{analysisMethod}. 

For obtaining the specific intensity function we use step size differences form previous specific intently values ($s_n$), described in Equation \ref{defineI_1}. We obtain the optimal value for each $s$ by using a random walk scheme. Starting from an initial guess of each $s$ (obtained from the wavelength averaged light curve), all $s$ are changed at once, with changes leading to better fit being used as new starting point for the next iteration. Converging on the optimal solution typically requires a large number of iterations, and for each possible combination of $s_n$ a unique intensity function is generated. In order to assess the quality of fit for this, we synthesize a transit light curve and then compare this to the observed data. When generating transit light curves we need the fraction of stellar light that is blocked by the planet, at each observed position of the planet. Computational time of the entire data analysis scales strongly with this disk integration, and we therefore use the following procedure to use computational time more effectively: The planet is divided into area segment as shown in Figure \ref{numSke}. The fraction of the stellar area that each of these segments cover is precalculated (as it is independent of the specific intensity function). The distance from the center of each area segment to the center of the stellar disk is also precalculated for each observed position of the planet (as this is also independent of the specific intensity function). As the specific intensity is calculated as a function of radial distance from the center of the star, we can simply interpolate this onto the (precalculated) distances to each area segment. This gives us the specific intensity at the center of each segment, and we approximate the specific intensity in the entire segment with this value. By multiplying this with the (precalculated) fraction of the stellar area that the segment covers and then summing over all segment, we get the intensity blocked by the entire planet. Thus disk integration is done in three simple steps: Interpolation, multiplication and summation. This allows for good accuracy at short computational time. The main advantage is amount of precalculation that this method allows for. The accuracy primarily scales with number of area segments the planet is divided into, which can easily be tweaked for each individual case to reach desired accuracy (larger planets requires more area segments).

\begin{figure}
	\centering
	\includegraphics[width=\hsize]{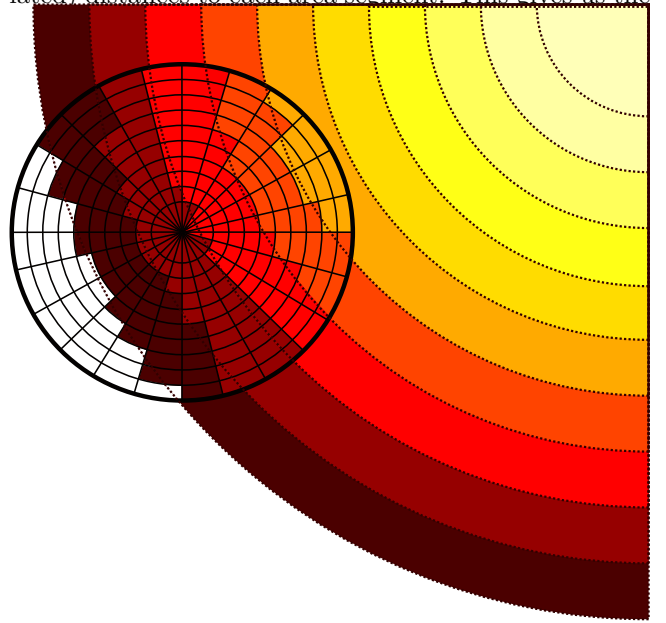}
	\caption{Sketch of planet disk integration scheme. The planet is divided into area segments (in our application we typically use segments, roughly 10 times, and planets are typically smaller in relation to the star). The specific intensity under each area segment is approximated by the value at the center. The combined contribution from the whole planet is calculated by first multiplying each area segment with the fraction of the stellar area it covers, and then summing over all segments. This allows for fast computations as neither the fraction of the stellar disk that is blocked by each segment nor the position of each segment on the stellar disk changes with specific intensity, and these can thus be precalculated and reused for creating light curves for any specific intensity function.}
	\label{numSke}
\end{figure}

\bibliography{mybiblE.bib}

\end{document}